\documentclass[oldversion]{aa}

\usepackage[varg]{txfonts}
\usepackage{siunitx}
\DeclareSIUnit\um{\micro\meter}
\DeclareSIUnit\Msun{M_{$\odot$}}

\usepackage{url}
\usepackage[tbtags,fleqn]{amsmath}
\usepackage{amsfonts}
\usepackage[x11names]{xcolor}
\usepackage{graphicx}
\usepackage{natbib}
\usepackage{aalongtable}

\newcommand{\cdbox}[1]{%
  \colorlet{currentcolor}{.}%
  {\color{DodgerBlue2}%
    \dbox{\color{currentcolor}#1}}%
}
\usepackage{dashbox,framed,color,ocg-p}
\newcommand{\ToggleLayer}[2]{%
  \leavevmode
  \pdfstartlink user {
    /Subtype /Link
    /Border [0 0 0]%
    /A <<
      /S/JavaScript
      /JS (
         var aOCGs = this.getOCGs(), Layer;
         var Layers = "#1".split(","), Active = -1, i, l;
         for (l=0; l<Layers.length; l++) {
           Layer = Layers[l];
           for (i=0; aOCGs && i<aOCGs.length; i++) {
             if (aOCGs[i].state && aOCGs[i].name == Layer) {
               Active = l;
               aOCGs[i].state = false;
             }
           }
           if (Active >= 0) break;
         }
         if (Active == -1) {
           for (l=0; l<Layers.length; l++) {
             if (Layers[l] == "") Active = l;
           }
         }
         Active = Active + 1;
         if (Active == Layers.length) Active = 0;
         Layer = Layers[Active];
         for (i=0; aOCGs && i<aOCGs.length; i++) {
           if (aOCGs[i].name == Layer) aOCGs[i].state = true;
         }
      )
    >>
  }#2%
  \pdfendlink
}

\newcommand{\e}{\mathrm e}

\newcommand{\diff}{\mathrm d}

\newcommand{\mincir}{\raise
  -2.truept\hbox{\rlap{\hbox{$\sim$}}\raise5.truept \hbox{$<$}\ }}
\newcommand{\magcir}{\raise
  -2.truept\hbox{\rlap{\hbox{$\sim$}}\raise5.truept \hbox{$>$}\ }}


\DeclareMathOperator{\Hs}{H}

\begin{document}

\title{\textit{Herschel-Planck} dust optical-depth and column-density maps}
\subtitle{I. Method description and results for Orion}
\titlerunning{\textit{Herschel-Planck} optical-depth maps -- I. Orion}
\author{Marco Lombardi\inst{1,4}, Herv\'e Bouy\inst{2}, Jo\~ao
  Alves\inst{3}, and Charles J.~Lada\inst{4}}
\mail{marco.lombardi@unimi.it} \institute{%
  University of Milan, Department of Physics, via Celoria 16, I-20133
  Milan, Italy \and Centro de Astrobiolog\'{\i}a, INTA-CSIC, PO Box
  78, 28691 Villanueva de la Ca\~nada, Madrid, Spain \and University
  of Vienna, T\"urkenschanzstrasse 17, 1180 Vienna, Austria \and
  Harvard-Smithsonian Center for Astrophysics, Mail Stop 72, 60 Garden
  Street, Cambridge, MA 02138} \date{Received ***date***; Accepted
  ***date***}

\abstract{%
  We present high-resolution, high dynamic range column-density and
  color-temperature maps of the Orion complex using a combination of
  \textit{Planck} dust-emission maps, \textit{Herschel} dust-emission
  maps, and 2MASS NIR dust-extinction maps.  The column-density maps
  combine the robustness of the 2MASS NIR extinction maps with the
  resolution and coverage of the \textit{Herschel} and \textit{Planck}
  dust-emission maps and constitute the highest dynamic range
  column-density maps ever constructed for the entire Orion complex,
  covering $\SI{0.01}{mag} < A_K < \SI{30}{mag}$, or
  $\SI{2D20}{cm^{-2}} < N < \SI{5D23}{cm^{-2}}$.  We determined the
  ratio of the \SI{2.2}{\um} extinction coefficient to the
  \SI{850}{\um} opacity and found that the values obtained for both
  Orion~A and B are significantly lower than the predictions of
  standard dust models, but agree with newer models that
  incorporate icy silicate-graphite conglomerates for the grain
  population.  We show that the cloud projected pdf, over a large
  range of column densities, can be well fitted by a simple power law.
  Moreover, we considered the local Schmidt-law for star formation, and
  confirm earlier results, showing that the protostar surface density
  $\Sigma_*$ follows a simple law $\Sigma_* \propto
  \Sigma_\mathrm{gas}^\beta$, with $\beta \sim 2$.}  \keywords{ISM:
  clouds, dust, extinction, ISM: structure, ISM: individual objects:
  Orion molecular cloud, Methods: data analysis}

\maketitle

\section{Introduction}
\label{sec:introduction}

Our inability to accurately map the distribution of gas inside a
molecular cloud has been a major impediment to understanding the star
formation process. This is because tracing mass in molecular clouds is
challenging when about 99\% of the mass of a cloud is in the form of
H$_2$ and helium, which are invisible to direct observation at the
cold temperatures that characterize these clouds. Tracing mass in
molecular clouds is currently achieved through use of column-density
tracers, such as molecular-line emission, thermal dust-emission, and
dust-extinction. The simplest and most straightforward of these by far
is dust-extinction, in particular, near-infrared (NIR)
dust-extinction, as it directly traces the dust opacity (without
assumptions on the dust temperature), and relies on the well-behaved
optical properties of dust grains in the NIR (e.g.,
\citealp{2013A&A...549A.135A}). The advantages of the NIR
dust-extinction technique as a column-density tracer have been
discussed independently by \citet{2009ApJ...692...91G}, who performed
an unbiased comparison between the three standard density-tracer
methods, namely, NIR dust-extinction (\textsc{Nicer},
\citealp{2001A&A...377.1023L}), dust thermal emission in the
millimeter and far-IR, and molecular-line emission.  These authors
found that dust-extinction is a more reliable column-density tracer
than molecular gas (CO), and that observations of dust-extinction
provide more robust measurements of column-density than observations
of dust-emission (because of the dependence of the latter on the
uncertain knowledge of dust temperatures, $T$, and dust emissivities,
$\beta$). This implies that in a massive star-forming cloud, where
cloud temperatures can vary significantly because of the large number
of embedded young stars and protostars, dust-emission maps are
fundamentally limited as tracers of cloud mass. This is particularly
true for the densest cloud regions where star formation takes place.

Although straightforward and robust, the NIR extinction technique is
nevertheless limited by the number of available background stars that
are detectable through a cloud \citep{1994ApJ...429..694L,
  1998ApJ...506..292A, 2001A&A...377.1023L, 2009A&A...493..735L}. This
implies that the resolution of a NIR extinction map is a function of
Galactic latitude and, to a minor extent, Galactic longitude. For
example, angular resolutions on the order of \SI{10}{arcsec} are
easily achievable toward the Galactic Bulge with modern NIR cameras on
\SI{10}{m} class telescopes (corresponding to a physical resolution of
the order of \SI{1000}{AU} for regions such as the Pipe Nebula,
Ophiuchus, Lupus, and Serpens).  But for regions of critical
importance for star formation, such as Orion~A, the cloud hosting the
nearest massive star formation region to Earth, only angular
resolutions of about \SI{1}{arcmin} (or physical resolutions on the
order of \SI{24000}{AU}) are currently achievable with similar
instrumentation, because of the location of this cloud toward the
anti-center of the Galaxy, and about $20^\circ$ off the Galactic
plane.

The recent public release of ESA's \textit{Planck} and
\textit{Herschel} thermal dust-emission data offers an excellent
opportunity to study entire giant molecular complexes away from the
Galactic plane, such as Orion, at resolutions on the order of 5000~AU,
or about five times better than what is currently possible with NIR
extinction techniques. The \textit{Planck} space observatory
\citep{2010A&A...520A...1T,2011A&A...536A...1P} is an ESA space
observatory launched on 14 May 2009 to measure the anisotropy of the
cosmic microwave background (CMB).  It observes the sky in nine
frequency bands covering \SIrange{30}{857}{GHz} with high sensitivity
and angular resolution from \SI{3}{arcmin} to \SI{5}{arcmin}.  Most
relevant to the study of thermal dust-emission from molecular clouds,
the High Frequency Instrument (HFI; \cite{2010A&A...520A...9L,
  2011A&A...536A...6P}) covers the \SIlist{100;143;
  217;353;545;857}{GHz} (or \SIlist{3000;2100;1400;850;550;350}{\um}
respectively) bands with bolometers cooled to \SI{0.1}{K}, providing a
large-scale view of entire molecular complexes with an unprecedented
sensitivity to dust-emission.  The \textit{Herschel} space observatory
\citep{2010A&A...518L...1P} is an ESA space observatory working in the
far-infrared and submillimeter bands.  The high sensitivity of
\textit{Herschel} imaging cameras PACS \citep{2010A&A...518L...2P} and
SPIRE \citep{2010A&A...518L...3G} are able to generate dust-emission
maps with dynamic ranges that are not possible from ground-based
bolometers, and reaching low column densities similar to those reached
by NIR dust-extinction, although with a uniform resolution across the
sky (of about \SI{12}{arcsec} at \SI{160}{\um}, \SI{18}{arcsec} at
\SI{250}{\um}, and \SI{36}{arcsec} at \SI{500}{\um}).

Unlike the \textit{Planck} satellite, however, \textit{Herschel} did
not observe the entire sky.  To maximize the number of clouds
observed, the strategy followed by the GTO teams was to map the
densest regions in the molecular clouds.  These observations provide
a unique high-resolution and high dynamic range view of the densest
star-forming structures, in particular, for clouds far from the
Galactic plane where the resolution of the NIR dust-extinction maps is
limited.  The obvious drawback of this choice is that the maps are
incomplete, missing the extended low-column-density regions containing
most of a cloud's mass, as seen in NIR extinction maps
\citep[e.g.][]{2006A&A...454..781L, 2010A&A...512A..67L,
  2008A&A...489..143L, 2011A&A...535A..16L} (Alves 2013, in prep.).

In this paper we present a high-resolution, high dynamic range
column-density map of the Orion complex using a combination of
\textit{Planck} dust-emission maps, \textit{Herschel} dust-emission
maps, and our own 2MASS NIR dust-extinction maps. The Orion
column-density maps presented in this paper combine the robustness of
the 2MASS NIR extinction maps with the resolution and coverage of the
\textit{Herschel} and \textit{Planck} dust-emission maps and
constitute the highest dynamic range column-density maps ever
constructed for the entire Orion complex, covering $\SI{0.05}{mag} <
A_K < \SI{10}{mag}$, or $\SI{1D21}{cm^{-2}} < N < \SI{2D23}{cm^{-2}}$.

The Orion star-forming region, being the most massive and most active
star-forming complex in the local neighborhood
\citep[e.g.][]{1986ApJ...303..375M, 1991psfe.conf..125B,
  1995A&A...300..903B, 2005A&A...430..523W, 2008hsf1.book..459B,
  2011A&A...535A..16L}, is probably the most often studied
molecular-cloud complex \citep[see][]{2008hsf1.book..459B,
  2008hsf1.book..483M,2013ApJS..207...10R, 2013ApJ...766L..17S}.  It
contains the nearest massive star-forming cluster to Earth, the
Trapezium cluster \citep[e.g.][]{1997AJ....113.1733H,
  2000AJ....120.3162L, 2002ApJ...573..366M, 2012ApJ...748...14D}, at a
distance of \SI{414}{pc} \citep{2007A&A...474..515M}).

This paper is organized as follows. In Sect.~2 we briefly describe the
data reduction process. Section~3 presents our approach to the problem
of converting dust-emission into column-density. Section~4 is devoted
to the application of the technique to the Orion~A and B molecular
clouds. We discuss the results obtained in Sect.~5. Finally, in Sect.~6
we present a summary.

We make use of PDF JavaScript to create figures with multiple layers:
this make it easier to perform direct comparisons between different
data or different results.  Figures with multiple layers have buttons
highlighted with a dashed blue contour in their captions.  The hidden
layers can be displayed only using a PDF reader with JavaScript
enabled, such as Adobe$\textsuperscript{\textregistered}$
Acrobat$\textsuperscript{\textregistered}$,
Foxit$\textsuperscript{\textregistered}$ Reader, or Evince.  We also
provide the hidden layers as separate figures in the appendix (in
the electronic form of the journal).

\section{Data reduction}
\label{sec:data-reduction}

The Orion molecular clouds were observed by the all-sky
\textit{Planck} observatory and by the \textit{Herschel Space
  Observatory} as part of the \textit{Gould Belt Survey}
\citep{2010A&A...518L.102A}.  We used the final data products of
\textit{Planck} \citep{2013arXiv1303.5062P}.  For \textit{Herschel}, a
first set of observations was obtained in parallel mode using the PACS
(at \SIlist{70;160}{\um}) and SPIRE (\SIlist{250;350;500}{\um})
instruments simultaneously.  An additional set was obtained using PACS
alone at \SI{100}{\um} in scan mode. Table~\ref{tab:1} gives an
overview of the observations.  More details about the observational
strategy can be found in \citet{2010A&A...518L.102A}.  The data were
pre-processed using the \textit{Herschel Interactive Processing
  Environment} \citep[HIPE][]{2010ASPC..434..139O} version 10.0.2843,
and the latest version of the calibration files.  The final maps were
subsequently produced using \textit{Scanamorphos version 21}
\citep{2012ascl.soft09012R}, using its \emph{galactic} option, which
is recommended to preserve large-scale extended emission.

\begin{table*}
\centering
\caption{\textit{Herschel} parallel mode and pointed observations
  used. \label{tab:1}}
\footnotesize
\begin{tabular}{lcSS[explicit-sign=+]ccS[group-minimum-digits=4,table-alignment=right,table-format=5]@{, }S[group-minimum-digits=4,table-alignment=left,table-format=5]}\hline\hline
Target name          &  Obs. ID       &  {R.A.} &  {Dec.}
&    Wavelengths ($\si{\um}$)    &  Obs. Date   &
\multicolumn{2}{l}{Exp. time (s)} \\
\hline
OrionA-C-1 &	1342204098/9	 &	84.66	&
-7.34 	& 70, 100, 160, 250, 350, 500 &	2010-09-06 &	7490 & 7197	 \\
OrionB-NN- &	1342205074/5	 &	88.37	&
2.59 	& 70, 100, 160, 250, 350, 500 &	2010-09-25 &	9866 & 10459	 \\
OrionA-S-1 &	1342205076/7	 &	85.66	&
-9.14 	& 70, 100, 160, 250, 350, 500 &	2010-09-26 &	12122 & 12349	 \\
OrionB-N-1 &	1342215982/3	 &	87.21	&
0.81 	& 70, 100, 160, 250, 350, 500 &	2011-03-13 &	7674 & 7735	 \\
OrionB-S-1 &	1342215984/5	 &	86.03	&
-1.75 	& 70, 100, 160, 250, 350, 500 &	2011-03-13 &	17094 & 17446	 \\
OrionA-N-1 &	1342218967/8	 &	83.46	&
-5.14 	& 70, 100, 160, 250, 350, 500 &	2011-04-09 &	14132 & 15567	 \\
\hline
\end{tabular}
\end{table*}

\section{Method}
\label{sec:method}

In this section we describe the procedure used to derive dust
(effective) temperature, optical-depth maps, and dust column-density
maps.  The method requires NIR extinction maps, for example derived
from the \textsc{Nicer} \citep{2001A&A...377.1023L} or \textsc{Nicest}
\citep{2009A&A...493..735L}, and far-infrared or submillimiter (FIR)
dust-emission maps (in our specific case, obtained from the
\textit{Planck and Herschel Space Observatories}) at different
wavelengths.

\subsection{Physical model}
\label{sec:physical-model}

Since molecular clouds are optically thin to dust-emission at the
frequencies and densities considered here, we describe the specific
intensity at a frequency $\nu$ as a modified blackbody:
\begin{equation}
  \label{eq:1}
  I_\nu = B_\nu(T) \bigl[ 1 - \e^{-\tau_\nu} \bigr] \simeq B_\nu(T)
  \tau_\nu \; ,
\end{equation}
where $\tau_\nu$ is the optical-depth at the frequency $\nu$ and
$B_\nu(T)$ is the blackbody function at the temperature $T$:
\begin{equation}
  \label{eq:2}
  B_\nu(T) = \frac{2 h \nu^3}{c^2} \frac{1}{\e^{h \nu / k T} - 1} \; .
\end{equation}
Following standard practice, we assumed that frequency dependence
of the optical depth $\tau_\nu$ can be written as
\begin{equation}
  \label{eq:3}
  \tau_\nu = \tau_{\nu_0} \left( \frac{\nu}{\nu_0} \right)^\beta \; ,
\end{equation}
where $\beta \simeq 2$ and where $\nu_0$ is an arbitrary reference
frequency.  Following the standard adopted by the \textit{Planck}
collaboration (see below Sect.~\ref{sec:absolute-fluxes}), we used
$\nu_0 = \SI{353}{GHz}$, corresponding to $\lambda = \SI{850}{\um}$,
and we indicate the corresponding optical depth as $\tau_{850}$.

The \textit{Herschel} bolometers respond to the in-beam flux density
$S_\nu$, that is, to the specific intensity integrated over the beam
profile.  Therefore, to convert the measured flux into an intensity,
we need to take into account the beam size at the specific wavelength.
As explained below (see Sect.~\ref{sec:sed-fit}), because of changes
of the beam size with frequency, in this step we need to choose
between two distinct models of dust-emission, pointlike or extended.

We stress that when using this physical model we are making the
assumption that temperature gradients are negligible along the line of
sight. This is of course an approximation, in particular at the low
temperatures that characterize molecular clouds where a small increase
of $T$ produces a large increase in the intensity.  Therefore, when
observing a cloud that has a gradient of temperature along the line of
sight, one will receive photons mostly from the warmer regions crossed
by the line of sight (typically, from the outskirts of dense regions).
Therefore, the temperature derived from a fit of the data with
Eq.~\eqref{eq:1} will not be a simple average of the dust temperatures
along the line of sight, but will be biased high
\citep{2009ApJ...696.2234S}; as a consequence, the optical depth will
be underestimated \citep{2011A&A...530A.101M}. These effects can be
very strong when large gradients are present, which is generally the
case toward embedded protostars that warm up their local environment,
or at a more extreme case, when a cluster of embedded ionizing massive
stars create an \textsc{Hii} region, as in the case of the Orion
Nebula. For these reasons, we interpret $T$ in Eq.~\eqref{eq:1} as an
\textit{effective dust temperature} for an observed dust column.

\subsection{SED fit}
\label{sec:sed-fit}

\begin{figure}[tp!]
  \centering
  \includegraphics[width=\hsize]{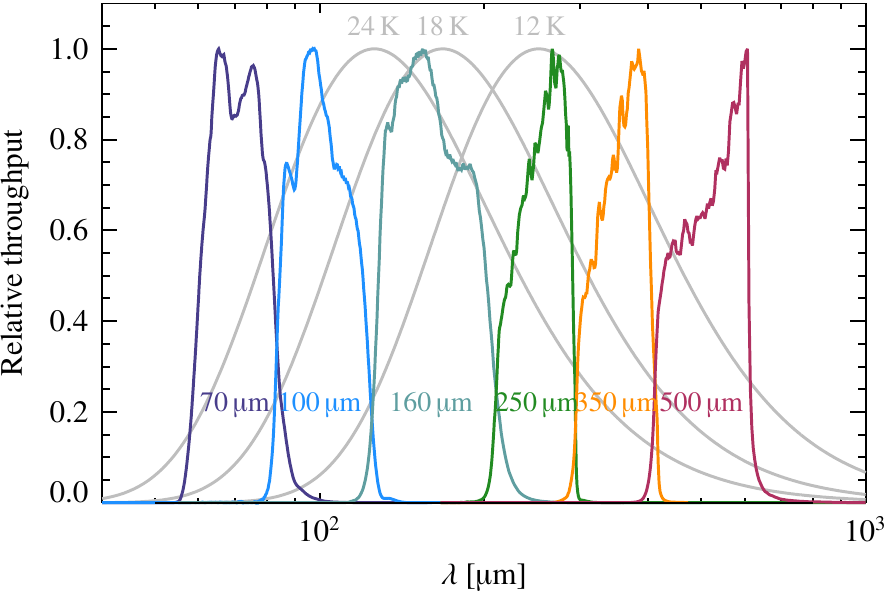}%
  \caption{Total throughputs of the PACS and SPIRE bands for extended
    emission and, superimposed as gray graphs, three modified
    blackbodies with $T \in \{\SI{12}{K}, \SI{18}{K}, \SI{24}{K}\}$
    and with $\beta = 1.8$. All lines are in arbitrary units (i.e.,
    the vertical axis only shows relative values).}
  \label{fig:1}
\end{figure}

If we know the optical depth $\tau_{850}$, the effective dust
temperature $T$, and the exponent $\beta$ in a given direction of the
sky, we can use Eqs.~(\ref{eq:1}--\ref{eq:3}) to infer the intensity
$I_\nu$ at each frequency $\nu$.  In reality, and if we aim to exploit
the higher resolution of \textit{Herschel}, we only have at our
disposal the fluxes measured by the PACS and SPIRE instruments at
specific wide bands.  For our purposes, it is useful to consider the
PACS \SIlist{100;160}{\um} bands, and the SPIRE
\SIlist{250;350;500}{\um} (the PACS \SI{70}{\um} band is not always
optically thin, and in many regions has a very low flux because it is
far away from the peak of the blackbody at the temperatures that
characterize molecular clouds, $\sim \SI{15}{K}$, see
Fig.~\ref{fig:1}).

To solve the inverse problem, that is, infer the optical depth and
effective dust temperature (and, eventually, the exponent $\beta$)
from the data, we proceeded as follows: we first convolved all
\textit{Herschel} data to the poorest resolution, that is, to
$\mathit{FWHM}_{\SI{500}{\um}} = \SI{36}{arcsec}$, corresponding to
the SPIRE \SI{500}{\um} data; then we performed a fit of the observed
spectral energy distribution (SED) by integrating the modified
blackbody intensity of Eq.~\eqref{eq:1} within each \textit{Herschel}
bandpass.  For the latter step we used the relative spectral response
functions (i.e., the total instrument throughputs) available for the
PACS and SPIRE bands, and for SPIRE, as recommended in the SPIRE user
manual, we corrected with the $\lambda^2$ factor corresponding to the
throughput for extended emission, which is appropriate for diffuse
emission (higher than the resolution of the instrument).\footnote{We
  deliberately ignore point sources in the analysis such as embedded
  protostars.  Therefore, in areas contaminated by these objects the
  derived dust column-density and temperature might not be accurate.
  For point sources one should use the original PSF without the
  $\lambda^2$ factor.}

\begin{figure*}
  \centering
  \includegraphics[width=\hsize]{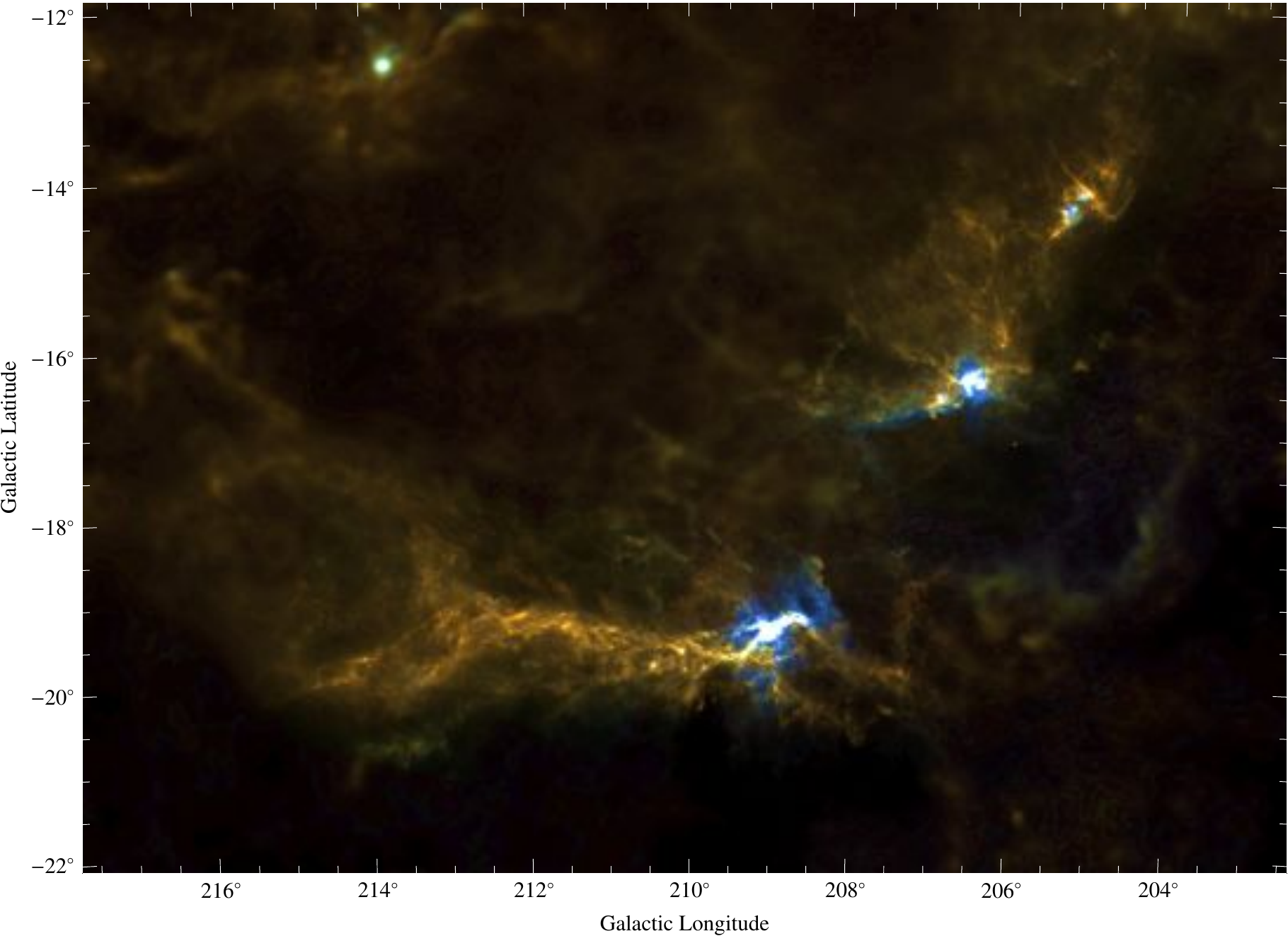}%
  \hspace{-\hsize}%
  \begin{ocg}{fig:2a}{fig:2a}{0}%
  \end{ocg}%
  \begin{ocg}{fig:2b}{fig:2b}{1}%
    \includegraphics[width=\hsize]{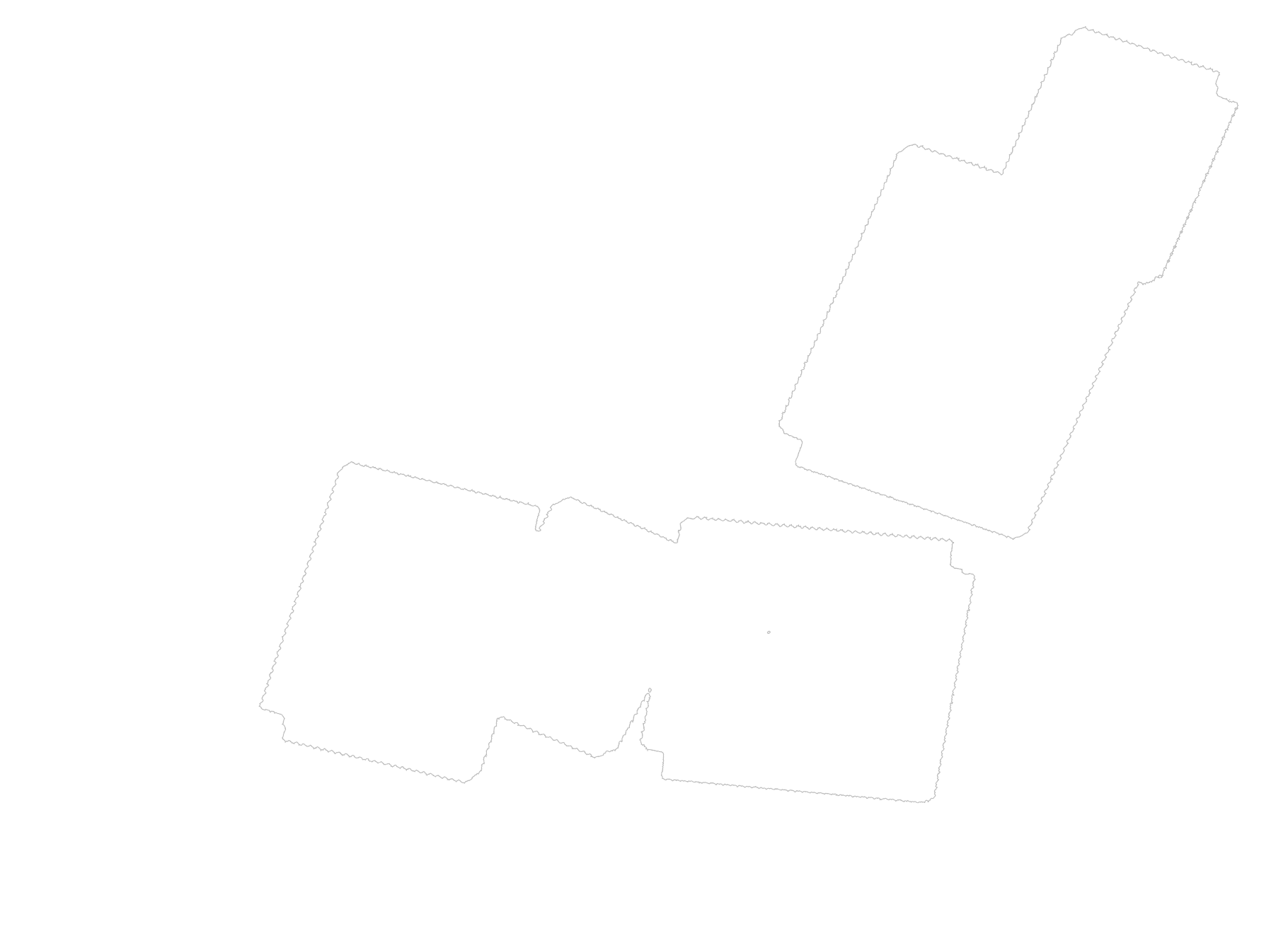}%
    \hspace{-\hsize}%
    \includegraphics[width=\hsize]{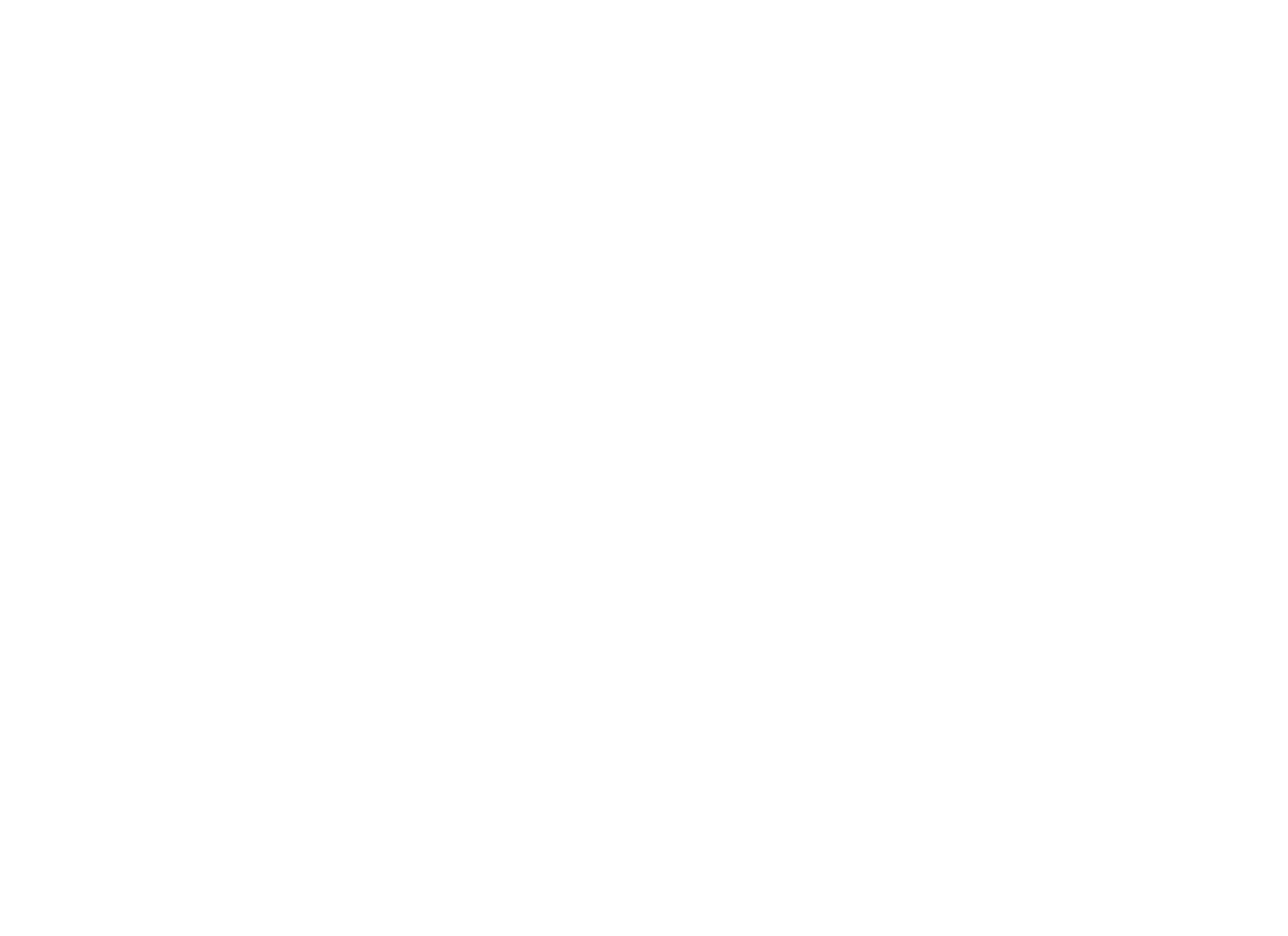}%
  \end{ocg}%
  \caption{Composite three-color image showing the
    \textit{Herschel}/SPIRE intensities for the region considered,
    where available (with the \SI{250}{\um}, \SI{350}{\um}, and
    \SI{500}{\um} bands shown in blue, green, and red).  For regions
    outside the \textsc{Herschel} coverage, we used the
    \textit{Planck/IRAS} dust model $(\tau_{850}, T, \beta)$ to
    predict the intensity that would be observed at the SPIRE
    passbands.  Note that the transition from \textit{Herschel} to
    \textit{Planck} is only visible because of the different
    resolution, and that otherwise there is no obvious discontinuity
    in the intensities.
    \ToggleLayer{fig:2b,fig:2a}{\protect\cdbox{Toggle labels}}}
  \label{fig:2}
\end{figure*}

The \textit{Herschel} bolometers measure the flux integrated within
each filter,
\begin{equation}
  \label{eq:4}
  \bar S = \frac{\int S_\mathrm{p,e}(\nu) R_\mathrm{p,e}(\nu) \, \diff
    \nu}{\int R_\mathrm{p,e}(\nu) \, \diff \nu} \, ,
\end{equation}
where $S_\mathrm{p,e}(\nu)$ is the in-beam source flux density and
$R_\mathrm{p,e}(\nu)$ is the specific passband throughput for point
(p) or extended (e) sources (cf.\ Fig.~\ref{fig:1}, where
$R_\mathrm{e}(\nu)$ is reported for the PACS and SPIRE passbands).
The \textit{Herschel} pipeline assumes that the source is point-like
and has an SED such that $S_\mathrm{p}(\nu) \nu = \mbox{constant}$
across the passband,
\begin{equation}
  \label{eq:5}
  S_\mathrm{p}(\nu) = S_\mathrm{p}(\nu_0) \frac{\nu_0}{\nu} \; .
\end{equation}
Therefore, the flux provided by the pipeline for each passband
corresponds to the flux that a point source with the spectral energy
distribution \eqref{eq:5} would have at the reference frequency
$\nu_0$, that is, $S_\mathrm{p}(\nu_0)$.  To obtain this quantity, the
pipeline converts the measured flux $\bar S$ into
$S_\mathrm{p}(\nu_0)$ by inserting Eq.~\eqref{eq:5} into
Eq.~\eqref{eq:4}.  This yields
\begin{equation}
  \label{eq:6}
  S_\mathrm{p}(\nu_0) = \left[ \frac{\int R_\mathrm{p}(\nu) \, \diff \nu}{\int
      (\nu_0/\nu) R_\mathrm{p}(\nu) \, \diff \nu} \right] \bar S
  \equiv K_\mathrm{4p} \bar S \; ,
\end{equation}
where, following the notation of the SPIRE observer manual, we have
called the correcting factor $K_\mathrm{4p}$.  In reality, the sources
of interest (dark clouds) present extended emission that follows a
modified blackbody SED.  If we consider the analogous definition of
$K_\mathrm{4e}$, that is,
\begin{equation}
  \label{eq:7}
  K_\mathrm{4e} \equiv \frac{\int R_\mathrm{e}(\nu) \, \diff \nu}{\int
      (\nu_0/\nu) R_\mathrm{e}(\nu) \, \diff \nu} \; ,
\end{equation}
we can write
\begin{equation}
  \label{eq:8}
  S_\mathrm{e}(\nu_0) = K_\mathrm{4e} \bar S =
  \frac{K_\mathrm{4e}}{K_\mathrm{4p}} S_\mathrm{p}(\nu_0) = \frac{
    \int S_\mathrm{e}(\nu) R_\mathrm{e}(\nu) \, \diff \nu}{\int (\nu_0 /
    \nu) R_\mathrm{e}(\nu) \, \diff \nu} \; ,
\end{equation}
where for the last step we have used Eq.~\eqref{eq:4}.

In summary, Eq.~\eqref{eq:8} provides a simple way to perform an SED
fit on the reduced \textit{Herschel} data:
\begin{itemize}
\item we first multiply for each SPIRE passband the flux reported by
  the pipeline, that is, $S_\mathrm{p}(\nu_0)$, by the correcting factor
  $C \equiv K_\mathrm{4e} / K_\mathrm{4p} = (0.9828, 0.9834, 0.9710)$
  for the $(250, 350, 500) \, \si{\um}$ bands, respectively,
\item we then perform an absolute flux calibration for the Herschel
  bands (see below Sect.~\ref{sec:absolute-fluxes}),
\item we assume a specific SED, compute the expected extended flux
  $S_\mathrm{e}(\nu_0)$ at each reference passband $\nu_0$ using the
  r.h.s.\ of Eq.~\eqref{eq:8}, and
\item finally, we modify the SED until we obtain a good match between
  the observed and theoretical fluxes.  For this step we use a simple
  $\chi^2$ minimization that takes into account the calibration errors
  (that we conservatively take to be equal to $15\%$ in all bands).
  Because of the degeneracies present in the $\chi^2$ minimization, we
  have kept $\beta$ fixed in the minimization, and fit only
  $\tau_{850}$ and $T$ to the data.  However, the spectral index
  $\beta$ was not kept constant across a cloud, but instead we used the
  local value of $\beta$ as estimated from the \textit{Planck}
  collaboration (see below Sect.~\ref{sec:absolute-fluxes}).
\end{itemize}

\subsection{Absolute fluxes}
\label{sec:absolute-fluxes}

Since the \textit{Herschel} SPIRE and PACS bolometers only provide
\textit{relative} photometry, we can only measure gradients of
intensities over the observed field.  This is usually no problem for
point source photometry, but represents a major difficulty for
obtaining photometry for extended emission.  Before performing the SED
fit described in the previous section, therefore, we performed an
\textit{absolute} calibration of all \textit{Herschel} passbands.

For this purpose, we used the maps released by the \textit{Planck}
collaboration \citep{2011A&A...536A...1P, 2011A&A...536A..19P,
  2013arXiv1303.5062P}. These maps report the results of an all-sky
SED fit for a modified blackbody [see Eq.  \eqref{eq:1}] using the
\textit{Planck}/HFI (\SI{350}{\um} to \SI{2}{mm}) and IRAS
(\SI{100}{\um}) data. The maps have an intrinsic resolution of
\SI{5}{arcmin} for the optical-depth and effective-dust temperature,
and \SI{35}{arcmin} for the spectral index $\beta$. [Note that the map
of the spectral index was also used to derive the local value of
$\beta$ that was used in the SED fit described in
Sect.~\ref{sec:sed-fit}.]

To obtain the absolute flux of each individual \textit{Herschel}
field, we proceeded as follows: from the \textit{Planck}
optical-depth, temperature, and spectral-index map we computed the
fluxes expected to be observed by \textit{Herschel} at the various
passbands.  For this step, we used the modified black-body model
\eqref{eq:1}, integrated over each passband as in Eq.~\eqref{eq:8}.
We then cross-correlated the \textit{Herschel} observations, degraded
to the \SI{5}{arcmin} resolution, with the computed expected fluxes,
and fitted a straight line to the fluxes,
\begin{equation}
  \label{eq:9}
  S_\mathrm{e}(\nu)_\mathrm{Herschel} = a_\nu + b_\nu \,
  S(\nu)_\mathrm{Planck} \; .
\end{equation}
The offset $a$ of the linear fit provides the absolute photometric
calibration of \textit{Herschel}. The slopes $b_\nu$ are always very
close to unity for all frequencies and all regions, which ensures that
our methodology is robust and that the \textit{Herschel} data
have a good relative photometry.  We repeated the same procedure for
each field and each passband separately.

Figure~\ref{fig:2} presents a color-composite image of the combined
reduced \textit{Herschel}/SPIRE data for the region considered here,
together with the predicted fluxes from \textit{Planck} at the three
SPIRE passbands. This figure graphically depicts the outcome of the
procedure described in this section, except that no convolution to the
\textit{Planck} resolution has been performed for the
\textit{Herschel} data (instead, the three SPIRE bands displayed in
the figure have been convolved to the \SI{500}{\um} resolution,
\SI{36}{arcsec}).  From this figure we can also appreciate the
temperature differences present in the clouds, which result in
different colors across the clouds.  Note also that the
high-temperature areas, which appear blue in the figure, are generally
associated with much brighter emission, a well-known effect of the
Planck law \eqref{eq:2}.  Moreover, the colors of the high-resolution
regions, observed by \textit{Herschel}, match the colors of the rest
of the field very well, where we used \textit{Planck} data: this is
an additional indication that the absolute calibration of the
\textit{Herschel} fluxes has been successful.

\subsection{Extinction conversion}
\label{sec:extinct-conv}

\begin{figure}[tp!]
  \centering
  \includegraphics[width=\hsize]{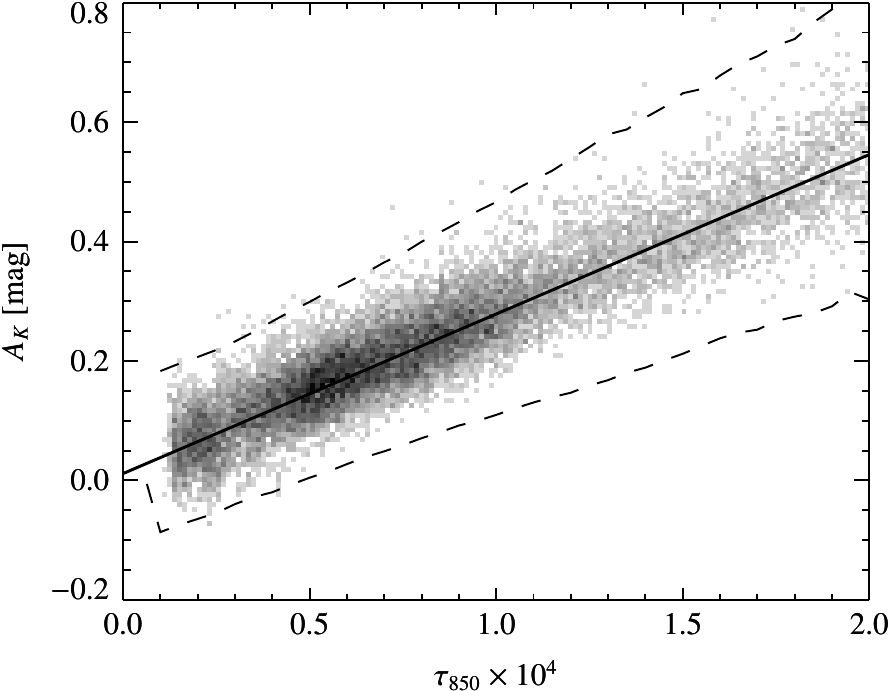}%
  \caption{Relationship between submillimiter optical-depth and
    NIR extinction in Orion~A.  The best linear fit, used to calibrate
    the data, is shown together with the expected 3-$\sigma$ region,
    as calculated from direct error propagation in the extinction
    map.}
  \label{fig:3}
\end{figure}

\begin{figure}[tp!]
  \centering
  \includegraphics[width=\hsize]{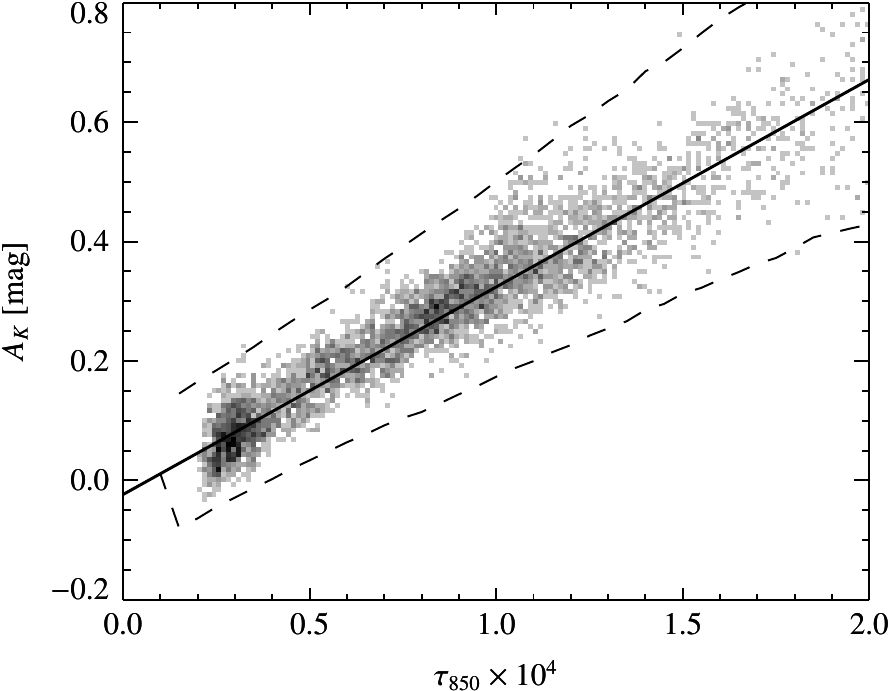}%
  \caption{Same as Fig.~\ref{fig:3} for Orion~B.}
  \label{fig:4}
\end{figure}

\begin{figure}[tp!]
  \centering
  \begin{ocg}{fig:5a}{fig:5a}{0}%
    \includegraphics[width=\hsize]{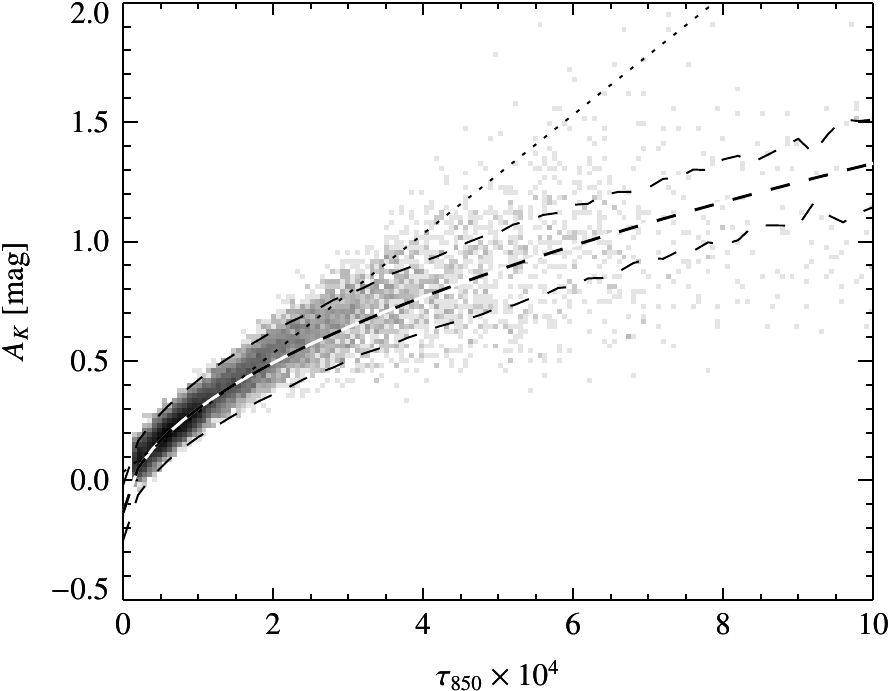}%
  \end{ocg}%
  \hspace{-\hsize}%
  \begin{ocg}{fig:5b}{fig:5b}{1}%
    \includegraphics[width=\hsize]{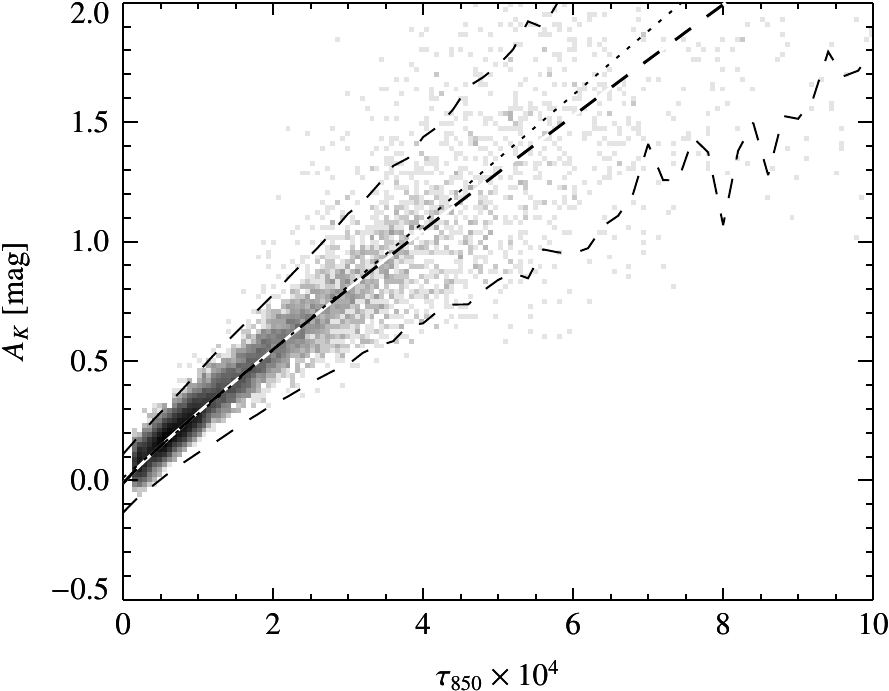}%
  \end{ocg}%
  \caption{Same as Fig.~\ref{fig:3}, but for a wider range of values.
    The plot shows hints of non-linearity for high values of dust
    column densities or optical-depths, as shown by the curved fit
    (dashed line).  We also report in this plot the linear fit
    obtained in the range of Fig.~\ref{fig:3} (dotted line).  A direct
    comparison between the results obtained from
    \ToggleLayer{fig:5a,fig:5b}{\protect\cdbox{\textsc{Nicest} and
        \textsc{Nicer}}} shows how important it is to account for
    unresolved substructures and foreground stars in molecular
    clouds.}
  \label{fig:5}
\end{figure}

\begin{figure}[tp!]
  \centering
  \begin{ocg}{fig:6a}{fig:6a}{0}%
    \includegraphics[width=\hsize]{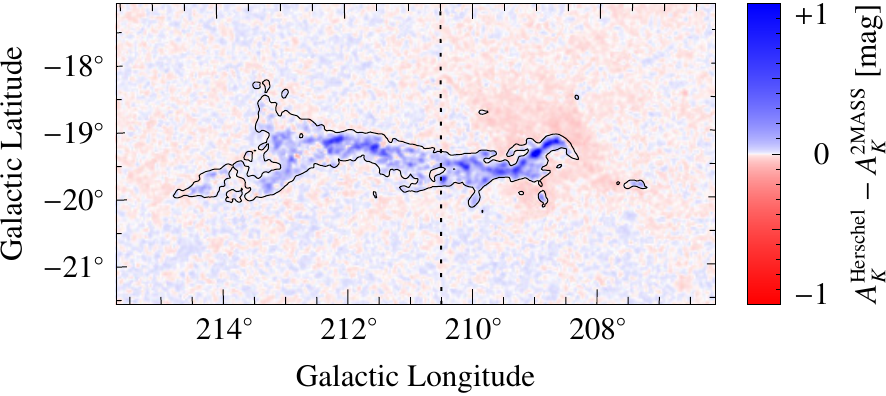}%
  \end{ocg}%
  \hspace{-\hsize}%
  \begin{ocg}{fig:6b}{fig:6b}{1}%
    \includegraphics[width=\hsize]{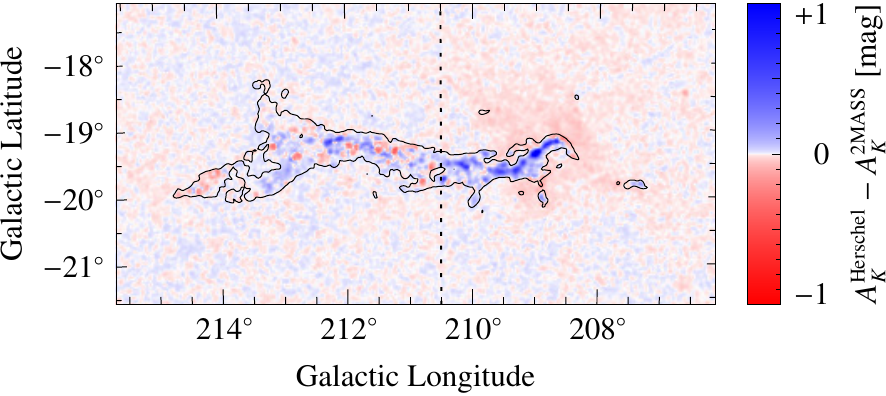}%
  \end{ocg}%
  \caption{Difference between the extinction predicted by
    \textit{Herschel}, using Eq.~\eqref{eq:11}, and the extinction
    measured with 2MASS/\textsc{Nicest}, with blue (red) indicating a
    positive (negative) difference.  The same figure for
    \textsc{Nicer} is available on a
    \ToggleLayer{fig:6a,fig:6b}{\protect\cdbox{different layer.}}}
  \label{fig:6}
\end{figure}

In principle, one could derive the slope of the linear relationship
between the optical-depth $\tau_{850}$ and the dust column by using a
suitable model of the interstellar grains:
\begin{equation}
  \label{eq:10}
  \tau_\nu = \kappa_\nu \Sigma_\mathrm{dust} \; ,
\end{equation}
where $\kappa_\nu$ is the opacity at the frequency $\nu$ and
$\Sigma_\mathrm{dust}$ is the dust-column density.  Determining the
dust opacity is a complicated task that requires a detailed knowledge
of the dust composition and properties \citep{1994A&A...291..943O},
which are often highly uncertain.  Since we have the NIR extinction
map at our disposal, we therefore preferred to derive the dust
column-density from $\tau_{850}$ by fitting a linear relationship
between $\tau_{850}$ and the $K$-band extinction $A_K$, obtained using
the \textsc{Nicest}/2MASS technique
\citep{2001A&A...377.1023L,2009A&A...493..735L,2011A&A...535A..16L},
after convolving all data to the same resolution (that is, the
resolution of the \textsc{Nicest} maps, $\mathit{FWHM} =
\SI{3}{arcmin}$):
\begin{equation}
  \label{eq:11}
  A_K = \gamma \tau_{850} + \delta \; .
\end{equation}
Note that the slope $\gamma$ is proportional to the ratio of
$\kappa_\mathrm{850}$, the opacity at \SI{850}{\um}, and of $C_{2.2}$,
the extinction coefficient at \SI{2.2}{\um}, since
\begin{equation}
  \label{eq:12}
  A_K = -2.5 \log_{10} \left( \frac{I_\mathrm{obs}}{I_\mathrm{true}}
  \right) = (2.5 \log_{10} \mathrm{e}) \, C_{2.2}
    \Sigma_\mathrm{dust} \; .
\end{equation}
Therefore we simply have $\gamma \simeq 1.0857 C_{2.2} /
\kappa_{850}$.  Conversely, we associate the coefficient $\delta$ to
either calibration inaccuracies (due, for example, to a control field
used in an extinction map that is not completely free of extinction,
or to an inaccurate photometric absolute calibration of the
\textit{Herschel} data), or to the presence of dust in the background
of the stars used to build the extinction map (that dust would clearly
escape extinction measurements, but would still be detected in
emission).  We found that a single fit within each cloud is
satisfactory, but different clouds require different fits.  In
Figs.~\ref{fig:3} and \ref{fig:4} we reports the result of these fits
for Orion~A and B, limiting the fit only to regions with
$\tau_{850} < \num{2D-4}$, where we empirically verified that
Eq.~\eqref{eq:11} is valid. The same figures also report the predicted
3-$\sigma$ boundaries around the fit of Eq.~\eqref{eq:11}, as
estimated from the statistical error on the extinction map alone (that
is, we ignored errors in the optical-depth).  The fact that the
datapoints are observed to lie within the marked region proves that
fit is very accurate and that the optical-depth map has a negligible
error at the resolution of the extinction map (that is,
\SI{3}{arcmin}).

In our specific case, we find $\delta_\mathrm{Orion\ A} =
\SI{0.012}{mag}$ and $\delta_\mathrm{Orion\ B} = \SI{-0.001}{mag}$,
while $\gamma_\mathrm{Orion\ A} = \SI{2640}{mag}$ and
$\gamma_\mathrm{Orion\ B} = \SI{3460}{mag}$.  It is leassuring that we
measure very low values for both offsets $\delta$; moreover, we do
not observe significant differences in the values of $\delta$ and
$\gamma$ within the tiles of a single molecular cloud, which would be
expected in case of calibration errors (we recall that the absolute
flux calibration was performed on each tile individually).  This can be
regarded as a success both of the calibrations for the extinction maps
(through a sensible choice for the control field as operated in
\citealp{2011A&A...535A..16L}) and for the \textit{Herschel} data
(through the use of the \textit{Planck} data).

As mentioned above, the coefficient $\gamma$ is simply linked to the
ratio of opacity at \SI{850}{\um} and of extinction at \SI{2.2}{\um}
(i.e., the sum of the opacity and scattering coefficient at
\SI{2.2}{\um}).  Differences in the values of $\gamma$, such as those
observed here, are probably to be related to differences in the dust
composition.  Differences in the opacity ratios are indeed common for
many similar studies carried out in the past: for example,
\citet{2003A&A...399.1073K} found that the opacity ratios determined
toward four cores of the IC~5146 span the range \num{1.9+-0.2e-4} to
\num{5.4+-0.3e-4}, which in terms of $\gamma$ would correspond to the
range \num{2000}--\num{5700} (see also \citealp{2011ApJ...728..143S}).
Using the data in Table~1 of \citet{1990ARA&A..28...37M}, we can
estimate the expected value of $\gamma$.  If we take $\beta
\simeq 1.8$ (value close to what was measured by \textit{Planck} in the
region we considered), we find
\begin{equation}
  \label{eq:13}
  \gamma \simeq 1.0857 \frac{C_{2.2}}{\kappa_{850}} =
  \frac{C_{2.2}}{\kappa_{250}} \left(
    \frac{\SI{850}{\um}}{\SI{250}{\um}} \right)^{-\beta} \simeq 2\,500
  \, ,
\end{equation}
which is very close to our measurements in Orion~A and not too far
from the value we obtain in Orion~B.  To reinforce this argument, we
also note that a recent analysis of the \textit{Planck} dust-emission
all-sky map \citep{2013arXiv1312.1300P} shows that the dust
optical-depth $\tau_{850}$ correlates well with the color excess of
quasars, with a relation $E(B-V) = \num{1.49+-0.03e4} \tau_{850}$,
which would imply $\gamma \simeq \num{4600}$ for $R \equiv A_V /
E(B-V) = 3.1$ and $A_K / A_V = 0.112$ \citep{1985ApJ...288..618R}.
However, as noted in \citet{2013arXiv1312.1300P}, the submillimiter
dust opacity can increase by a factor 3 in high-density regions, which
would cause $\gamma$ to decrease by a similar factor.  In summary, the
results for $\gamma$ are perfectly within the currently accepted range
of expected values.  Instead, our finding seems to be in conflict with
the emission and absorption coefficients computed by
\citet{2001ApJ...548..296W}, which would predict (for their size
distribution ``B'' with $R = 5.5$) $\gamma \simeq \num{4800}$ (and
even higher values for lower values of $R$).  Since $\gamma$ is
directly connected to the grain composition and size distribution, the
discrepancy we find might indicate that the \citet{2001ApJ...548..296W}
models may need to be revised, at least to explain the dust
properties in Orion~A and B.

To better understand this problem, we also considered the
\citet{1994A&A...291..943O} models.  These models provide dust
opacities under various conditions, but unfortunately \textit{not}
dust scattering cross-sections.  Since extinction is the sum of
opacity and scattering, we cannot apply these models directly.  To
obtain some estimates, however, we assumed the albedo, that is, the
ratio between opacity and extinction cross-sections, to be $0.5$ at
\SI{2.2}{\um}, and 1 at \SI{850}{\um} (in other words, we assumed that
opacity has the same cross section of scattering at \SI{2.2}{\um}, and
that scattering is negligible at \SI{850}{\um}).  These values,
although somewhat arbitrary, agree with the
\citet{2001ApJ...548..296W} models.  In our conditions, it seems
appropriate to use a \citet{1977ApJ...217..425M} dust distribution
with thin ice mantles.  Then, for a coagulation time of
\SI{1e5}{years} and a density of \SI{1e6}{cm^{-3}} we find $\gamma
\simeq \num{5100}$ (the predicted value of $\gamma$ decreases to
\num{3700} for models with a density of \SI{1e8}{cm^{-3}}, but this
value of course seems inappropriate for giant molecular complexes).

The newer \citet{2011A&A...532A..43O} models\footnote{See also
  \url{http://astro.berkeley.edu/~ormel/software.html}.} seem instead
to be able to reproduce the observed values for $\gamma$.  These
authors provided a list of opacities and extinction coefficients for a
few aggregate models at various coagulation times.  For our clouds, it
seems reasonable to use coagulation times on the order of 1 to
\SI{3}{Myr}.  With this choice, we find very reasonable values for
$\gamma$: for example, for the fully ice-coated aggregate
\texttt{(ic-sil, ic-gra)} (referred to as ``Type-II'' mixing in
\citealp{2011A&A...532A..43O}, and consisting of silicates and
graphite grains with ice manthles mixed within the aggregates) the
predicted value for $\gamma$ is $\sim \num{2580}$, very close to our
observations for Orion~A.  On the other hand, if we instead consider
the \texttt{ic-sil+gra} model (i.e., a spatial mixture of aggregates
consisting of either ice-coated silicate or graphite materials,
referred to as ``Type-I'' mixing), we find $\gamma \simeq \num{3800}$,
not too far from what was observed in Orion~B.  In summary, although
the dust models depends on quite a few parameters (dust composition,
mixture type, presence of ice mantles, grain size distribution,
coagulation time), we find it possible to accomodate the observed
values of gammas within the range of reasonable models.

Across a wider range of opacities (typically, for $\tau_{850} >
\num{2e-4}$) the relationship between NIR extinction and optical-depth
is no longer linear.  A good fit is obtained with the empirical
relation
\begin{equation}
  \label{eq:14}
  A_K = c_1 + c_2 \tau_{850}^{c_3} \; .
\end{equation}
Interestingly, for Orion~A we are able to recover an almost perfect
linearity if we remove the region around the Trapezium from the
analysis, that is, for $l > \SI{210.5}{\degree}$ (see
Fig.~\ref{fig:5}).  This suggests that the non-linearity is a
consequence of including regions for which the 2MASS/\textsc{Nicest}
extinctions or the \textit{Herschel} gray-body SED fit are inaccurate
(or both).  To better understand this problem, we produced a map that
shows the differences between the extinction, as inferred from the
optical-depth $\tau_{850}$ used in Eq.~\eqref{eq:11}, and the
2MASS/\textsc{Nicest} extinction (Fig.~\ref{fig:6}).  Note that the
blue regions, that is, those where the \textit{Herschel} column-density
exceeds the 2MASS extinction, are mostly confined to the region around
the Trapezium and closely follow the locations of known embedded
clusters in the cloud.  This suggests that the extinction map in these
regions is biased low as a result of the contamination from embedded
stars (observed therefore through a lower column-density than
genuine background stars; see \citealp{2009A&A...493..735L} and
\citealp{2005A&A...438..169L} for deeper discussions of these matters).
We therefore conclude that the extinction provided by
\textit{Herschel} is more reliable in these regions.  The same map
also shows an extended light-red area around the ONC.  This area
correlates very well with the hot regions in Fig.~\ref{fig:9}, and
therefore we suspect that there the \textit{Herschel} column-density
is underestimated.  As a likely explanation, we mention temperature
gradients along the line of sight (which are probable in regions
characterized by a temperature much higher than average),
which would induce an underestimate of $\tau_{850}$ (see
Sect.~\ref{sec:physical-model}).  Interestingly, when \textsc{Nicer}
is used, we see that the blue regions tend to fill the entire cloud,
indicating that the \textsc{Nicer} extinction map is biased low in all
regions with high extinction.

We stress that the test just performed is a strong confirmation of the
reliability of extinction studies in regions that are not contaminated by
embedded clusters and contain sufficient background stars.  Our
analysis shows in particular that the extinction measured by
\textsc{Nicest} is fully consistent with the completely independent
analysis carried out using the \textit{Herschel} data.  Note, however,
that a technique such as \textsc{Nicer}, which is optimized but does
not take into account the effect of foreground stars or small-scale
inhomogeneities, is clearly biased and cannot be used to probe
high-column density regions.  As discussed in \citet{2009A&A...493..735L},
this kind of bias is expected in basically all extinction techniques
(with the mentioned exception of \textsc{Nicest}), but this is the
first time we are in the position of proving its existence in real
data.

Given the results of this section, in the following use only the
coefficient $\gamma$ in the conversion from optical-depth to
extinction, that is, we set $A_K = \gamma \tau_{850}$.  Doing so,
we ignore $\delta$ because we consider the small measured offsets
of Eq.~\eqref{eq:11} as biases present in the extinction measurements.

\subsection{Higher resolution optical-depth maps}
\label{sec:high-resol-opac}

We have already mentioned the non-trivial interpretation of the
effective dust-temperature.  A posteriori, however, one can verify
that the derived maps of effective-dust temperature in most cases
appear to be significantly smoother than the optical-depth maps (see
also below).  This observation suggests that we can compute the term
$B_\nu(T)$ of Eq.~\eqref{eq:1} using a low-resolution map, and
evaluate simply as $\tau_\nu = I_\nu / B_\nu(T)$.

In practice, we applied this technique by using the temperature maps
obtained from the modified blackbody fit of the \textit{Herschel}
bands (and therefore computed at the lower resolution, corresponding
to $\mathit{FWHM}_{\SI{500}{\um}} \simeq \SI{36}{arcsec}$) together
with the SPIRE250 intensity maps, which are characterized by
$\mathit{FWHM}_{\SI{250}{\um}} \simeq \SI{18}{arcsec}$. Hence, this
technique allowed us to improve the resolution of our optical-depth
maps by a factor two.  Note that this technique is the simplest one to
derive higher resolution maps from dust-emission data at different
resolutions \citep{2013A&A...553A.113J}.  Other options are available
\citep[see][]{2013A&A...557A..73J}, but the gain obtained is limited,
and this is obtained at the price of a significantly more complicated
implementation (in particular, the most promising technique, ``method
E'' of \citealp{2013A&A...557A..73J} cannot be easily used over large
regions).

In a sense, the use of a temperature map at the \SI{36}{arcsec}
resolution in an optical-depth SED fit at \SI{18}{arcsec} is similar
to the use of the \textit{Planck}-fitted spectral index $\beta$ in the
SED fit at \SI{36}{arcsec} resolution.  It is nevertheless important
to verify the effects of this choice in the final high-resolution
optical-depth map.  Toward this goal we evaluated the
\textit{relative} variation of the term $B_\nu(T)$ for the passband
\SI{250}{\um}, that is, the multiplicative term in the SED that is
temperature dependent, and we verified that it is on the order of
$\sim 30 \%$.  In contrast, the relative change in optical-depth is
approximately a factor 10 larger. Hence, the changes in the observed
flux at \SI{250}{\um} are clearly dominated by optical-depth gradients
and not by temperature gradients, and this justifies the
implementation and use of higher resolution maps.

\begin{figure*}
  \centering
  \begin{ocg}{fig:7a}{fig:7a}{0}%
    \includegraphics[width=\hsize]{Orion/fig02}%
  \end{ocg}%
  \hspace{-\hsize}%
  \begin{ocg}{fig:7b}{fig:7b}{1}%
    \includegraphics[width=\hsize]{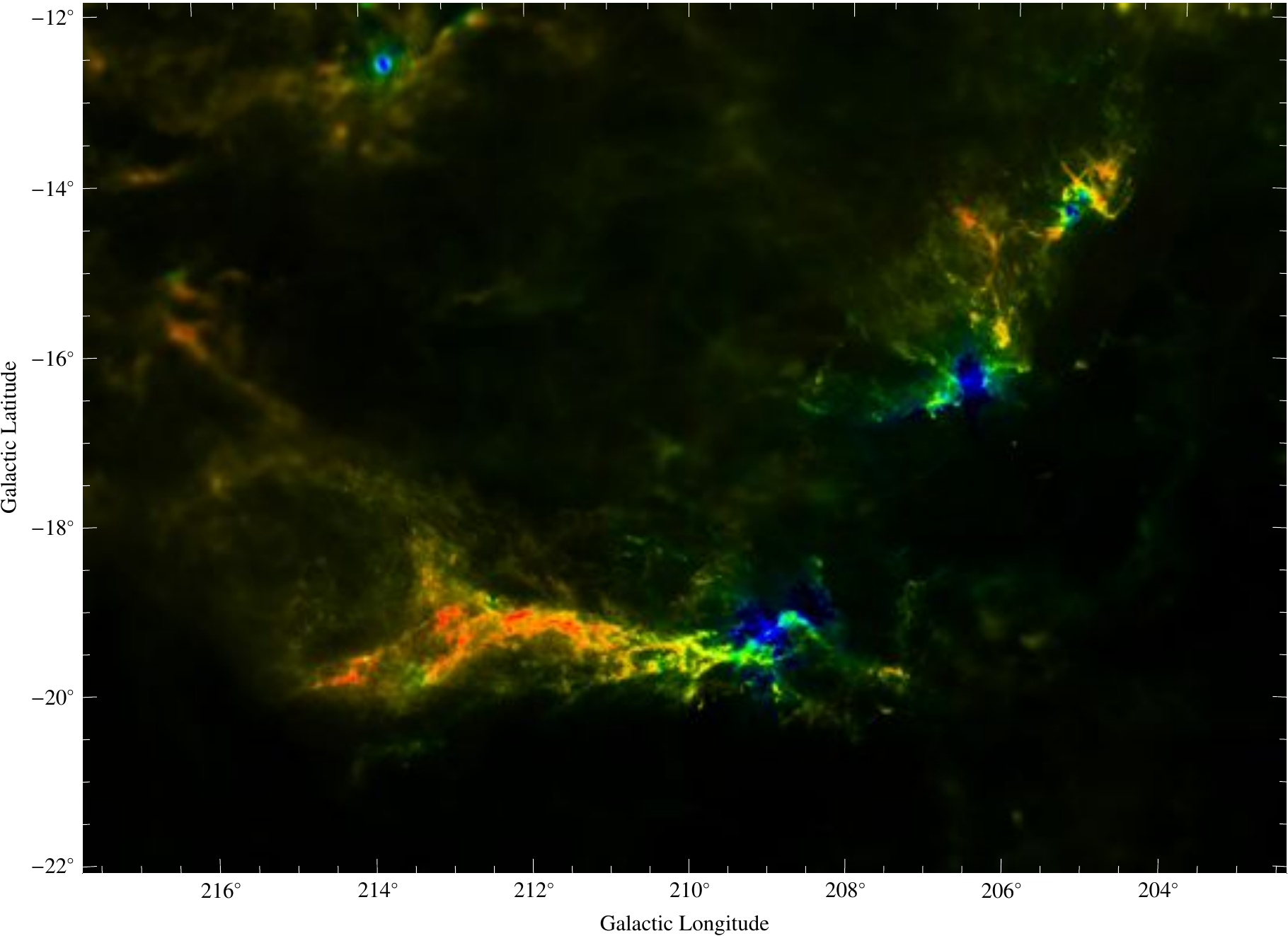}%
  \end{ocg}
  \hspace{-\hsize}%
  \begin{ocg}{fig:7c}{fig:7c}{0}%
  \end{ocg}%
  \begin{ocg}{fig:7d}{fig:7d}{1}%
    \includegraphics[width=\hsize]{Orion/fig02b}%
    \hspace{-\hsize}%
    \includegraphics[width=\hsize]{Orion/fig02_newlab}%
  \end{ocg}%
  \caption{Combined optical depth-temperature map for Orion~A and
    B.  The image shows the optical-depth as intensity and the
    temperature as hue, with red (blue) corresponding to
    low temperatures (high temperatures).  By comparing this image
    with the one shown in
    \ToggleLayer{fig:7b,fig:7a}{\protect\cdbox{Figure~1}} one can
    appreciate that regions with relatively high temperatures emit
    much higher fluxes even if the optical-depth is substantially
    lower.  \ToggleLayer{fig:7d,fig:7c}{\protect\cdbox{Toggle
        labels}}}
  \label{fig:7}
\end{figure*}

\begin{figure*}
  \centering
  \begin{ocg}{fig:8a}{fig:8a}{0}%
    \includegraphics[width=\hsize]{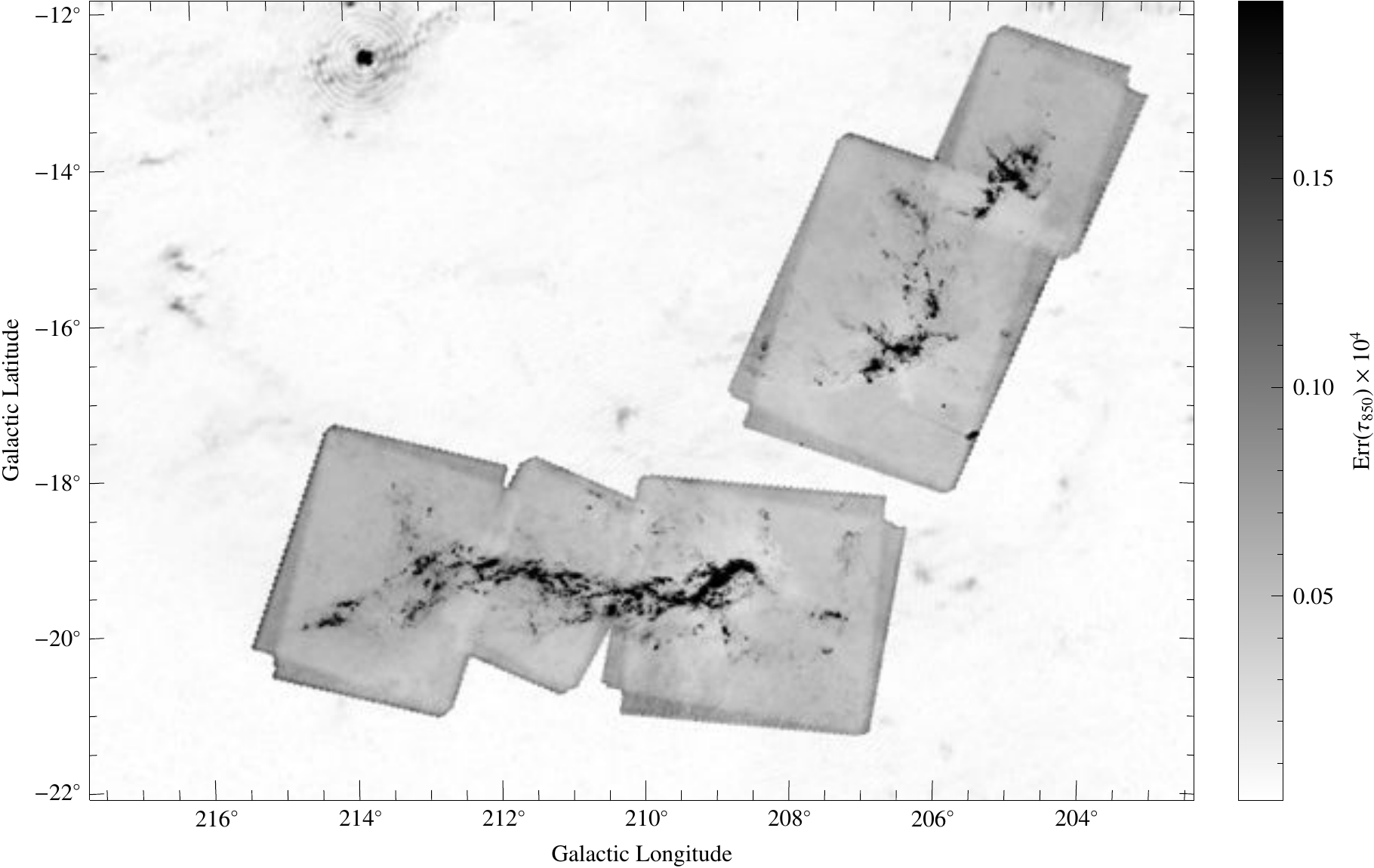}%
  \end{ocg}%
  \hspace{-\hsize}%
  \begin{ocg}{fig:8b}{fig:8b}{1}%
    {\color{white}\rule{\hsize}{0.63\hsize}}%
    \hspace{-\hsize}%
    \includegraphics[width=\hsize]{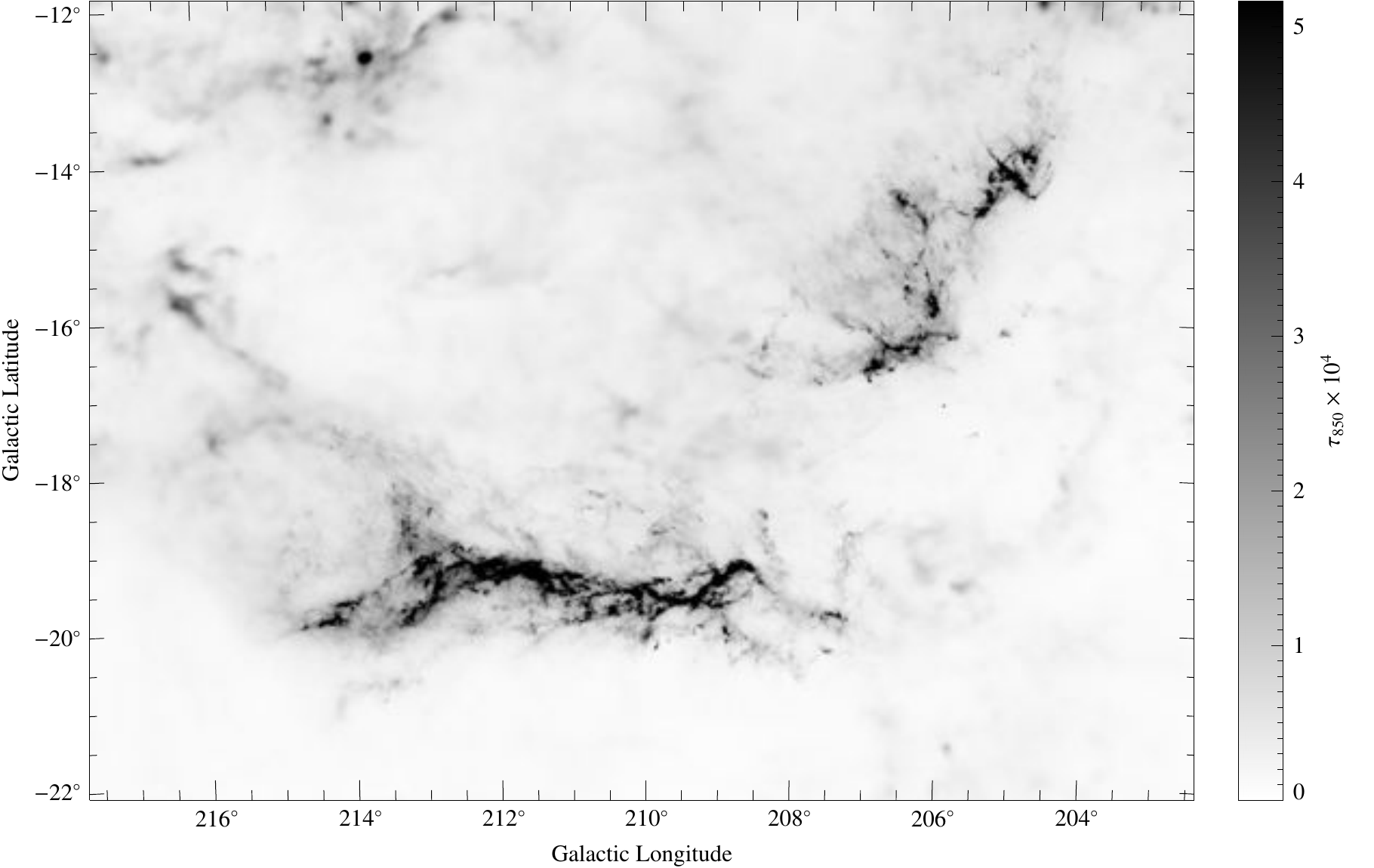}%
  \end{ocg}%
  \caption{Optical-depth map for the field and, on a
    \ToggleLayer{fig:8a,fig:8b}{\protect\cdbox{different
        layer}}, the corresponding error map.  The error-map image
    clearly shows the areas where the \textit{Herschel} data are
    available.  The resolution of the image varies from \SI{5}{arcmin}
    (corresponding to the \textit{Planck} data) to \SI{36}{arcsec}
    (for the \textit{Herschel}-covered areas).}
  \label{fig:8}
\end{figure*}

\begin{figure*}
  \centering
  \begin{ocg}{fig:9a}{fig:9a}{0}%
    \includegraphics[width=\hsize]{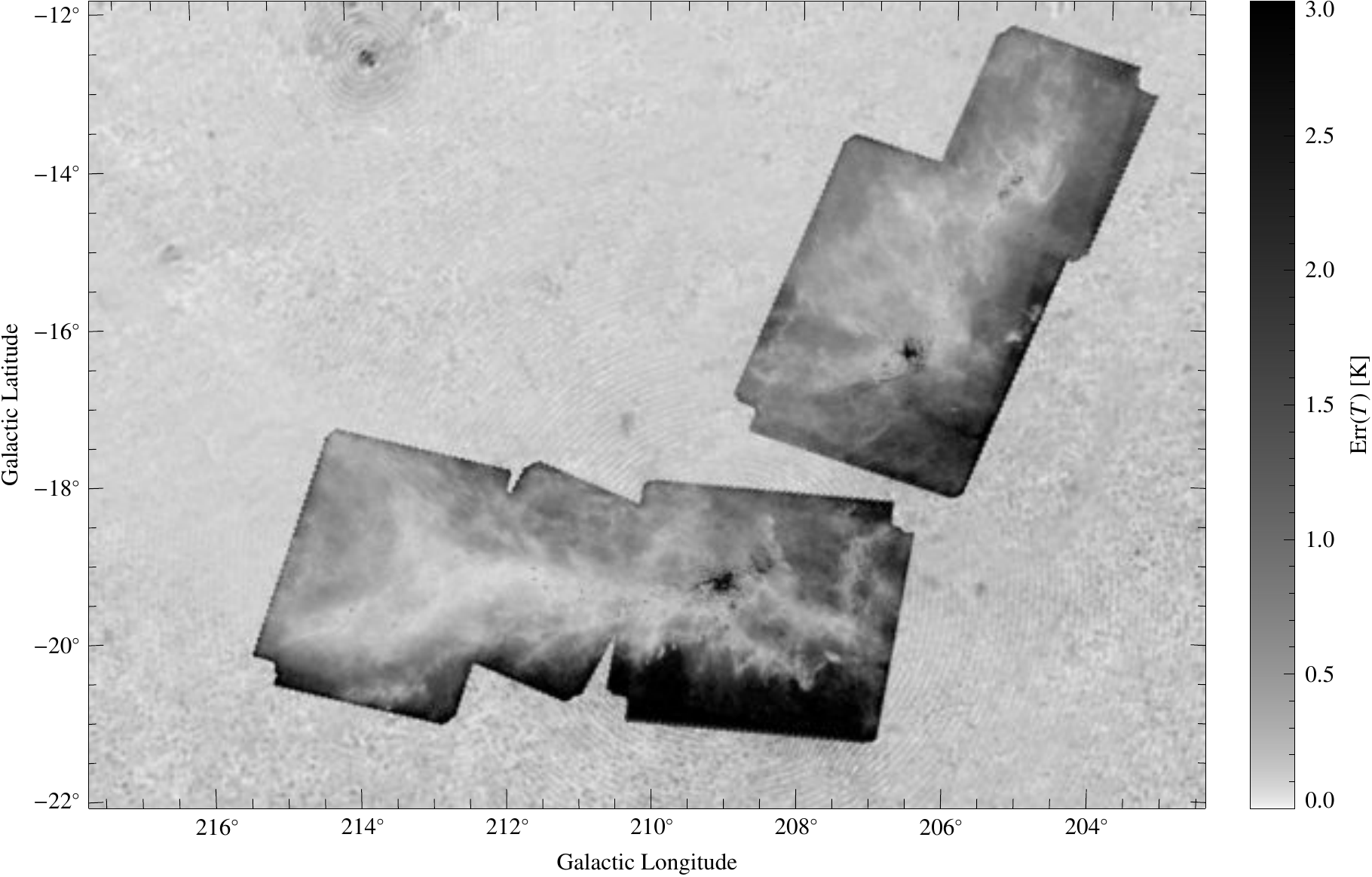}%
  \end{ocg}%
  \hspace{-\hsize}%
  \begin{ocg}{fig:9b}{fig:9b}{1}%
    {\color{white}\rule{\hsize}{0.639\hsize}}%
    \hspace{-\hsize}%
    \includegraphics[width=\hsize]{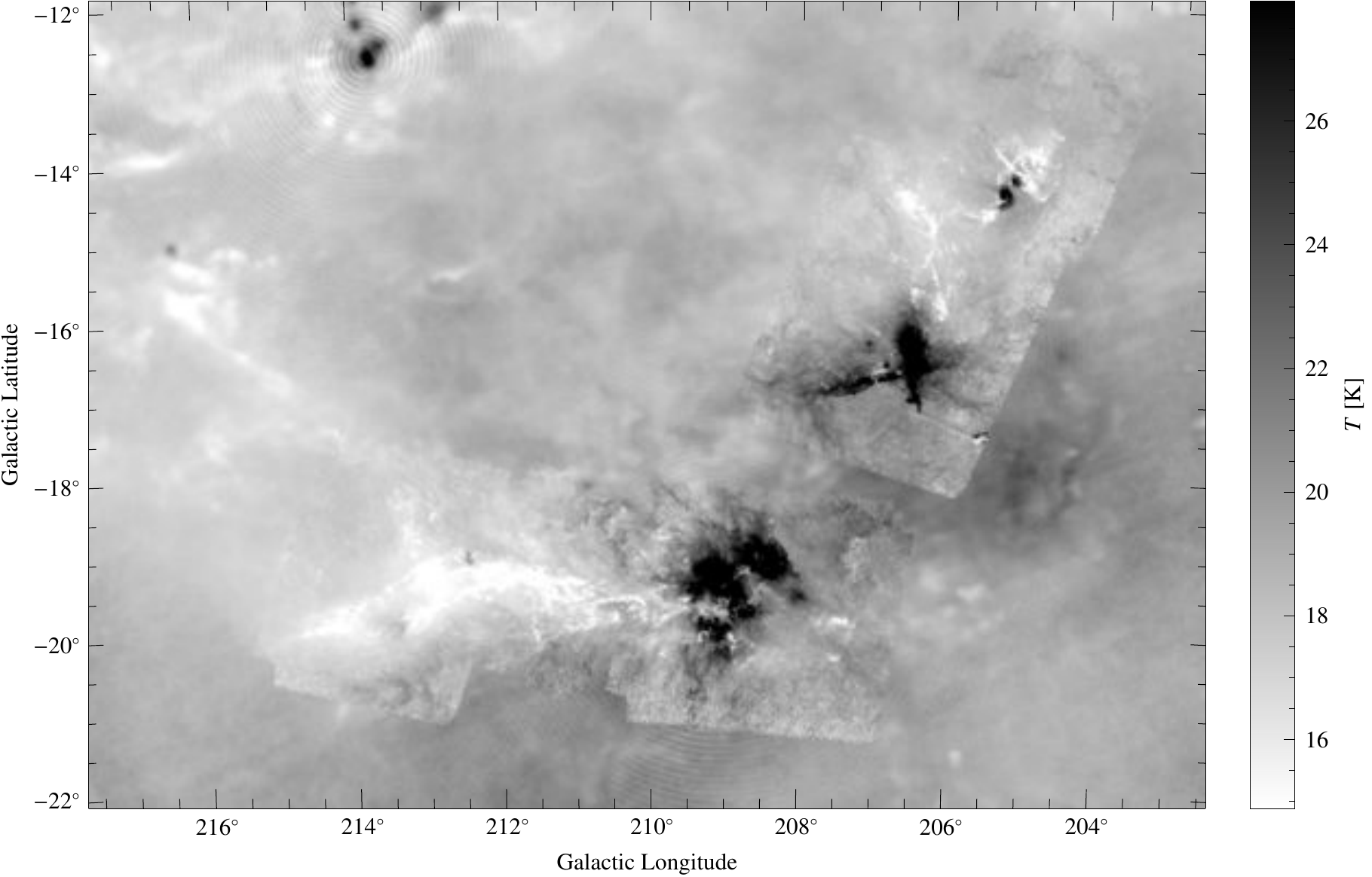}%
  \end{ocg}%
  \caption{Effective dust-temperature map for the field and, on a
    \ToggleLayer{fig:9a,fig:9b}{\protect\cdbox{different layer}},
    the corresponding error map.  Similarly to Fig.~\ref{fig:8}, the
    error-map image shows the areas where the \textit{Herschel} data
    are available and the resolution varies from \SI{5}{arcmin}
    (corresponding to the \textit{Planck} data) to \SI{36}{arcsec}
    (for the \textit{Herschel}-covered areas).}
  \label{fig:9}
\end{figure*}

\subsection{Implementation}
\label{sec:implementation}

A specific C code was written to perform various steps of the
pipeline, and in particular the SED fit.  The code is fully parallel
and takes advantages of the linear dependences of the models on
parameters [for example, the linear dependence of the modified
blackbody model \eqref{eq:1} on the optical-depth $\tau_{850}$]; all
this allowed us to analyze large regions very quickly.  The nonlinear
chi-square minimization was performed using the C-version of the
MINPACK-1 least squares fitting
library\footnote{\url{http://www.physics.wisc.edu/~craigm/idl/cmpfit.html}}.

In a typical pipeline run we perform the following steps:
\begin{enumerate}
\item We perform a standard reduction of \textit{Herschel} data and
  multiply the SPIRE data by the $C$ correcting factors.  At the same
  time, we produce an extinction map in the same area using data from
  the 2MASS-PSC archive, and also retrieve the optical-depth,
  temperature, and spectral-index maps created by the \textit{Planck}
  collaboration.
\item We convolve the \textit{Herschel} reduced images to reach a
  \SI{5}{arcmin} resolution, and we warp and re-grid the images to
  match the \textit{Planck} data projection.
\item We generate from the \textit{Planck} data the expected fluxes at
  \textit{Herschel} passbands, and we compare this pixel by pixel.  By
  performing the linear fit of Eq.~\eqref{eq:9} we recover the offset
  $a$ of each \textit{Herschel} waveband and verify that the linear
  coefficient $b$ is close to unity.
\item We return to the reduced \textit{Herschel} images (whose fluxes
  are now absolutely calibrated) and convolve them to the same
  resolution (typically, the \SI{36}{arcsec} resolution of the SPIRE
  \SI{500}{\um} data).
\item We perform an SED fit pixel by pixel using a modified blackbody
  as a model, leaving the optical-depth and effective-dust temperature
  as free parameters; in contrast, the local value of the spectral
  index $\beta$ is taken from the \textit{Planck}/\textit{IRAS} fit.
\item Finally, we also build a higher resolution map from the SPIRE
  \SI{250}{\um} band by inferring the optical-depth from the observed
  flux (and assuming the $\beta$ from the \textit{Planck} and $T$ from
  the \SI{36}{arcsec} resolution SED fit).
\end{enumerate}

\section{Results}
\label{sec:results}

\subsection{Optical-depth and temperature maps}
\label{sec:opac-temp-maps}

\begin{figure}[tp!]
  \centering
  \includegraphics[width=\hsize]{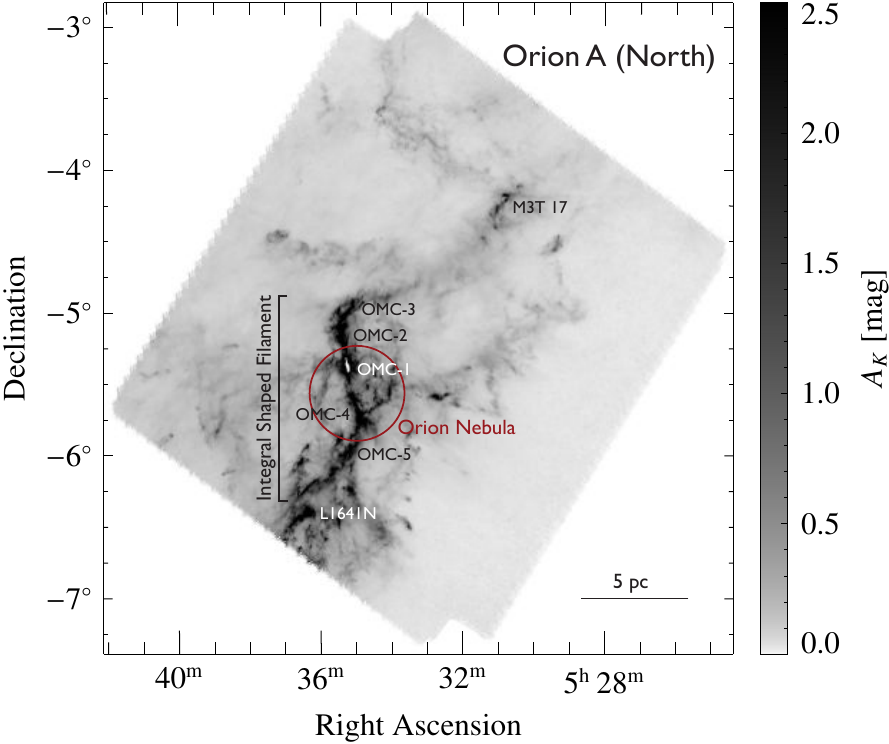}%
  \caption{Optical-depth map of Orion A central at a resolution of
    \SI{18}{arcsec}, converted into units of $A_K$ from
    cross-correlation with the 2MASS/\textsc{Nicest} map.}
  \label{fig:10}
\end{figure}

\begin{figure*}
  \centering
  \includegraphics[height=0.49\hsize]{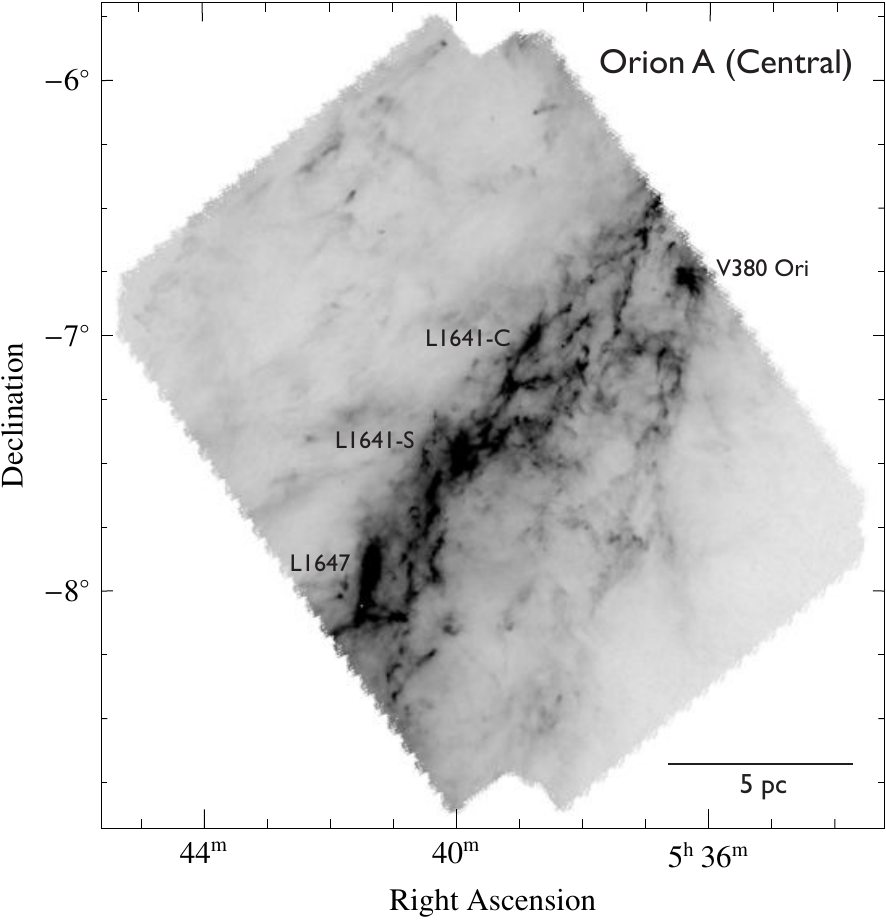}%
  \hfill
  \includegraphics[height=0.49\hsize]{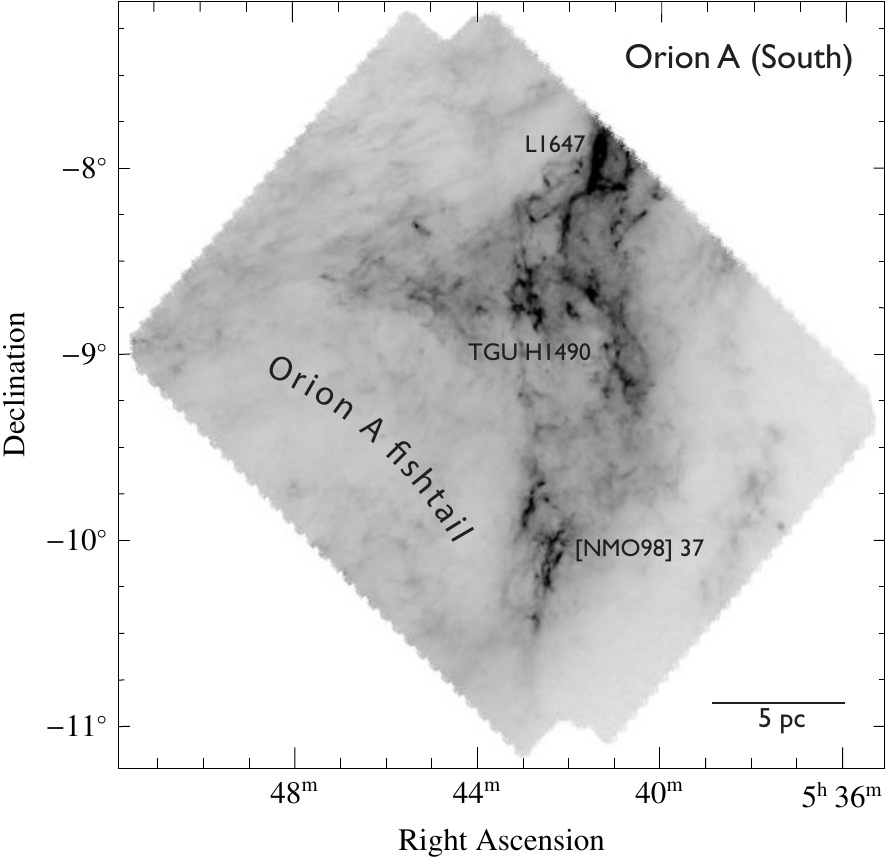}\\
  \includegraphics[height=0.49\hsize]{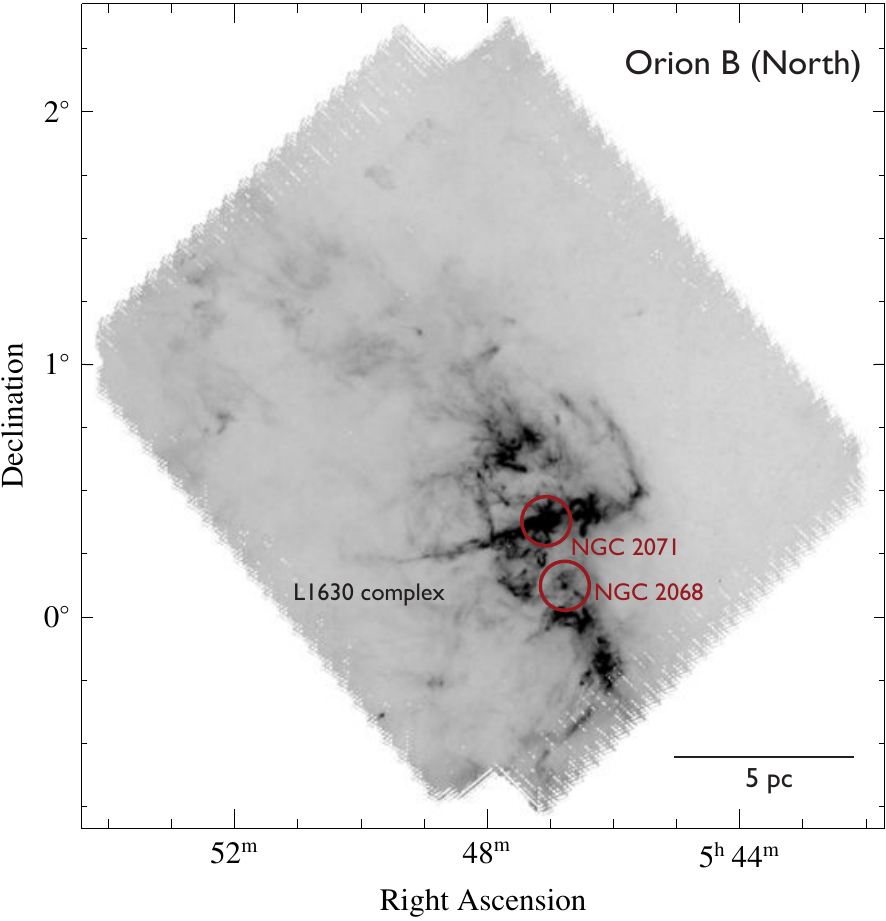}%
  \hfill
  \includegraphics[height=0.49\hsize]{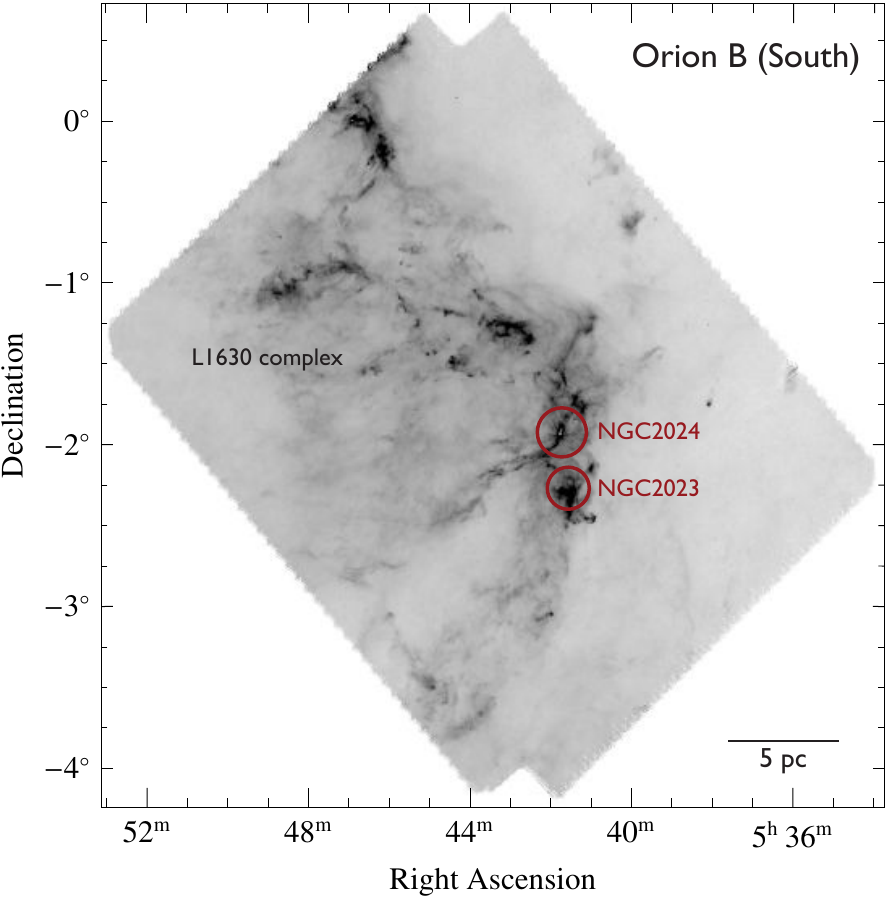}%
  \caption{Same as Fig.~\ref{fig:10} for Orion A north, Orion A south,
    Orion B north, and Orion B NN.}
  \label{fig:11}
\end{figure*}

The main final products of our custom pipeline are the optical-depth
and temperature maps of the region.  To exploit the unique sensitivity
and the large-scale view of the entire molecular cloud complex
provided by \textit{Planck}, we combined the \textit{Herschel} map
with the \textit{Planck}/\textit{IRAS} data, even though there is a
large difference in resolution between these data (\SI{36}{arcsec}
vs.\ \SI{5}{arcmin}).

Figure~\ref{fig:7} shows the combined optical-depth-temperature map:
the effective dust temperature is represented using different values
of hue (from red for $T \le \SI{12}{K}$ to blue for $T \ge
\SI{30}{K}$), while the intensity is proportional to the optical
depth.  This representation has the advantage of showing all the main
final products in a single image and of suppressing the relatively
large uncertainties on the effective dust temperature present in the
\textit{Herschel} tiles when the amount of dust, and thus the
optical-depth, are low.  It is also interesting to compare
Fig.~\ref{fig:7} with Fig.~\ref{fig:2} and directly appreciate how
relatively hot regions in the maps, typically associated with hot
early-type stars, are in Fig.~\ref{fig:2}.

Figure~\ref{fig:8} individually shows the optical-depth measured in
the whole field by either \textit{Herschel} or
\textit{Planck}/\textit{IRAS}.  On a different layer, the figure also
reports the associated errors, which are typically around \num{5e-6}
for the \textit{Herschel} data.  The \textit{Planck} data have
significantly smaller errors (mostly because of the much longer
exposure time), but of course have a much poorer resolution.  The
error map is also useful to visually reveal the exact shapes of the
areas covered by \textit{Herschel}.

The corresponding effective-dust temperature map is shown in
Fig.~\ref{fig:9}, together with its error (on a different layer).  It
is interesting to observe a number of features of this image.  First,
the specific area considered shows a relatively wide range of
temperatures, in particularly related to OB stars present in the
massive star-forming regions (the Orion Nebula Cluster, NGC~2024 in
Orion~B, and the Mon~R2 cluster).  It is also evident that the
temperature drops in dense regions of the cloud (provided there are
no early-type stars present in the region): this is particularly
evident in the ``spine'' of the Orion~A molecular cloud.

The error on the temperature map has a wide range as well.  The
largest error on the temperature is observed within the
\textit{Herschel} boundaries, but outside the densest regions of the
cloud (in particular, to the west of Orion~A, where it reaches values
of \SIrange{3}{4}{K}).  Instead, errors on the effective dust
temperature in regions where the optical-depth is high can be as low
as $\sim \SI{0.1}{K}$, similar to lower than the errors observed in
the \textit{Planck} region.  Regions at the boundaries of the
\textit{Herschel} coverage and with large errors on the effective dust
temperature are often associated with mismatches with the
\textit{Planck} data.  These systematic mismatches are due to a
combination of the low signal-to-noise ratio and of the different
wavelength coverage of the maps.

Finally, we note that the \textit{Planck} temperature maps show clear
ringing around bright areas.  These artifacts are a result of the use
of different wavelengths (with slightly different resolutions) in
nonlinear algorithms in the \textit{Planck} pipeline.

Figures~\ref{fig:10} and \ref{fig:11} show the optical-depth maps of
several regions of Orion~A and B at the \SI{18}{arcsec} resolution.
These maps have been obtained using the technique described in
Sect.~\ref{sec:high-resol-opac} (we prefer to show individual regions
in separate figures to better show the level of detail achieved). Note
that for these higher resolution maps we do not report any error
estimate, since assessing the error is non-trivial. We can isolate
three main sources of errors, however: the statistical, photometric
error on the SPIRE250 flux (which directly propagates to an error in
the final optical-depth); the statistical error on the temperature
(which propagates to the optical-depth through the Planck function
$B_\nu(T)$ at $\lambda = \SI{250}{\um}$, and which is also correlated
to the error on the SPIRE250 flux); and the error on the temperature
due to the different resolution (which of course is not statistical in
nature and thus difficult to quantify).

\subsection{Validation}
\label{sec:validation}

\begin{figure}[tp!]
  \centering
  \includegraphics[width=\hsize]{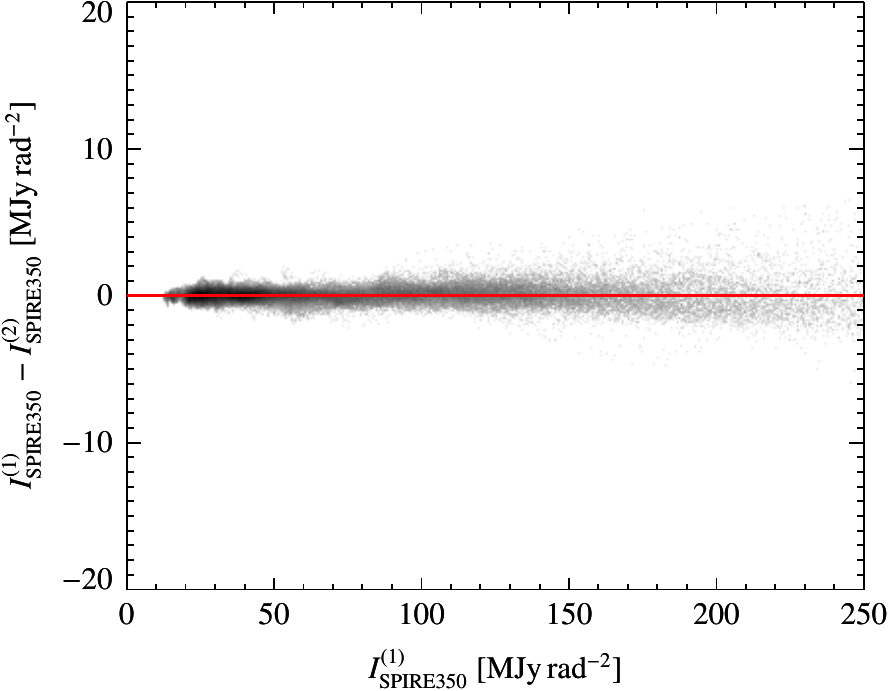}%
  \caption{Relationship between the SPIRE~350
    fluxes as measured on Orion~A southern and central fields (which
    are partially overlapping), \textit{after} absolute flux
    calibration.  The median of the flux difference between the two
    fields, $\sim \SI{56}{kJy.rad^{-2}}$ is approximately 6 times
    lower than the average noise level, and more than 1\,000 times
    lower than the median flux level in the overlapping area.}
  \label{fig:12}
\end{figure}

\begin{figure}[tp!]
  \centering
  \includegraphics[width=\hsize]{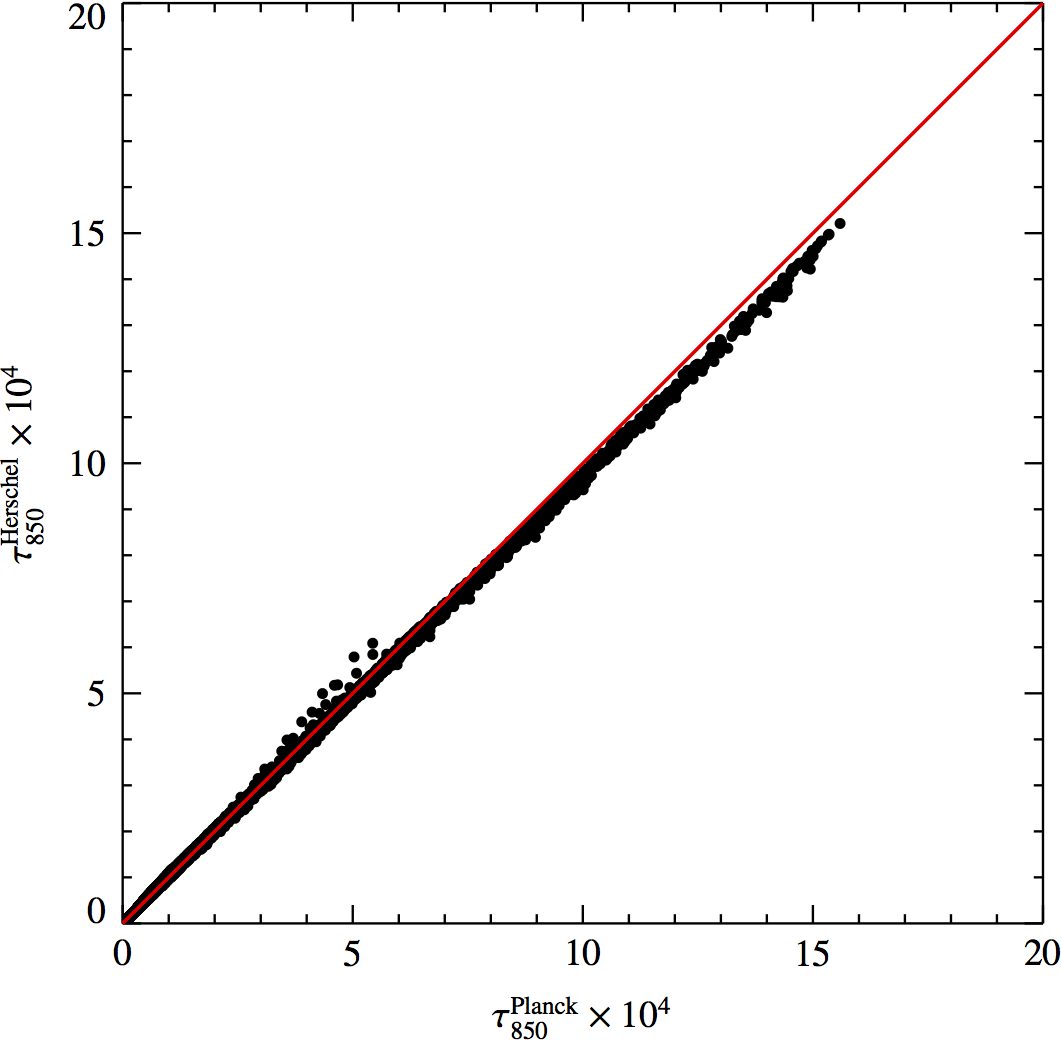}%
  \caption{Comparison between the optical-depth derived from the
    \textit{Planck} data and that derived from our calibration of
    the \textit{Herschel} data in Orion~A, after convolving them at
    the resolution of \textit{Planck} (\SI{5}{arcmin}).  The excellent
    agreement confirms that the \textit{Herschel} data are perfectly
    calibrated to \textit{Planck} and that the derived optical-depth
    does not show any systematic effect.}
  \label{fig:13}
\end{figure}

\begin{figure}[tp!]
  \centering
  \includegraphics[width=\hsize]{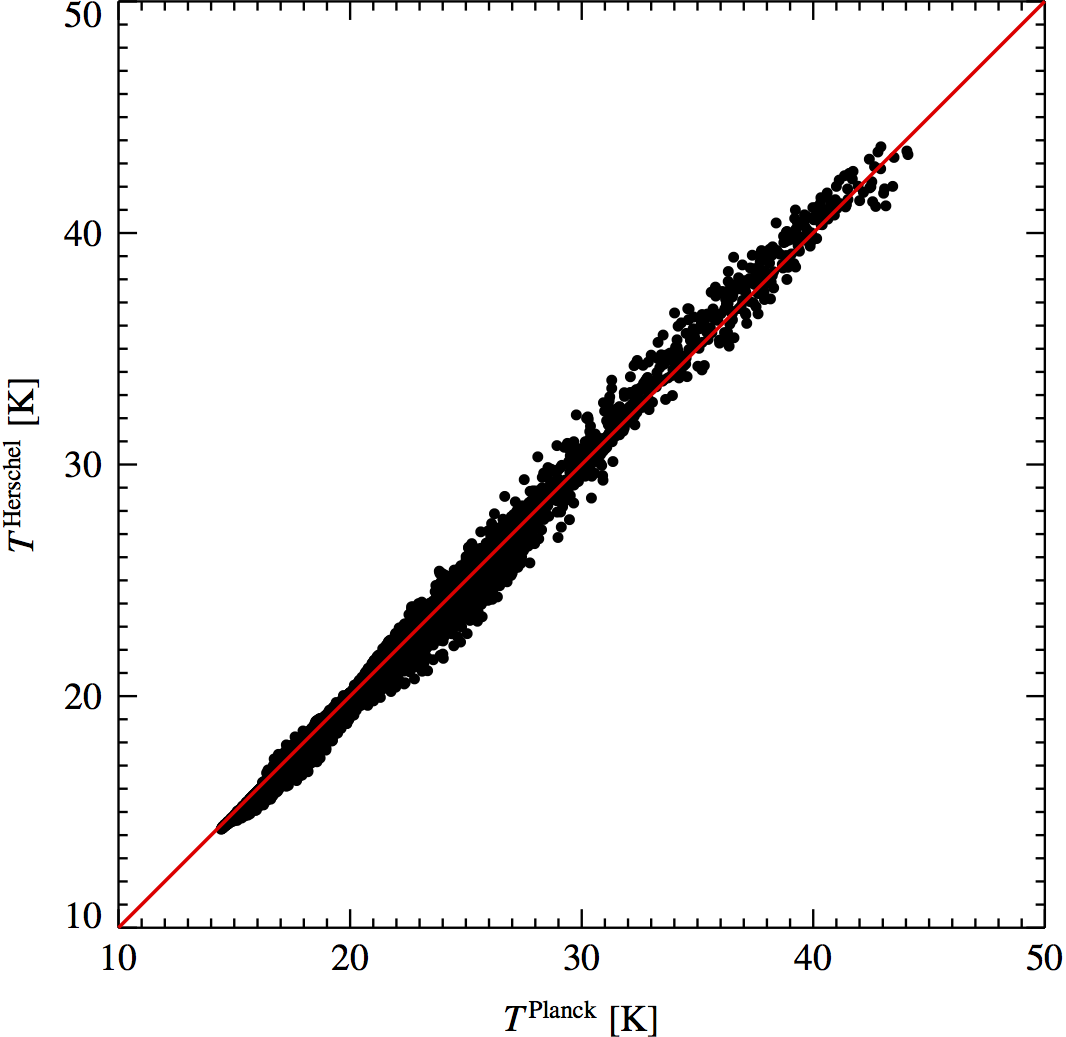}%
  \caption{Comparison between the effective dust temperature derived
    from the \textit{Planck} data and that derived from our
    calibration of the \textit{Herschel} data in Orion~B, after
    convolving them at the resolution of \textit{Planck}
    (\SI{5}{arcmin}).}
  \label{fig:14}
\end{figure}

It is obvious that in general the results obtained appear to be
reliable: the shapes of the clouds are the ones we know from
independent measurements, the ranges of effective dust temperatures
measured are reasonable, and the fact that the effective dust
temperature drops in the dense regions is consistent with our
expectations.  However, we clearly need to and can go beyond these
simple qualitative aspects to assess the reliability of our
data. Indeed, throughout the analysis we have performed a series of
tests to cross-validate the results. In this section, we intend to
summarize these tests (some of which were mentioned in
Sect.~\ref{sec:method}) and to present new ones.

One of the critical aspects of calibrating \textit{Herschel} data is
the removal of the offsets in the individual bands and fields, in
other words, the conversion from relative to absolute fluxes.  As
explained, this is achieved by direct comparison with the predicted
fluxes from the optical-depth, temperature, and spectral index from
the \textit{Planck} maps.  We can check this step in two different
ways during the calibration: (1) as mentioned above, the calibration
slope, that is, parameter $b$ of Eq.~\eqref{eq:9}, must be close to
unity and (2) in overlapping areas of independent \textit{Herschel}
observations, the fluxes measured must agree.  We performed both
checks, and the results of one of the second tests, the comparison of
the SPIRE 250 fluxes in the Orion south and central fields, are reported
in Fig.~\ref{fig:12}.

Figures \ref{fig:3} and \ref{fig:14} provide another indirect check of
the consistency of the data. These plots show the relationship between
the submillimiter optical-depth and the extinction, or equivalently,
the ratio of extinction at \SI{2.2}{\um} and opacity at \SI{850}{\um},
$C_{2.2}/\kappa_{850}$ in the range $\tau_{850} < \num{2e-4}$.
Since both the submillimeter optical-depth and the extinction are
directly proportional to the dust column-density, we expect to see a
linear relation, and indeed this is what we observe (up to high
extinctions, provided \textsc{Nicest} is used).  Additionally, the
scatter observed around the linear fit is fully consistent with the
error in the extinction measurements (which, we recall, are based on
relatively shallow 2MASS data).

A similar check can be carried out by comparing the optical depth as
derived from the \textit{Herschel} data with that obtained by the
\textit{Planck} team. To carry out this test we first convolved all
\textit{Herschel} bands to \SI{5}{arcmin} resolution and then
performed an SED fit on the convolved data. The results of this
process are shown in Figs.~\ref{fig:13} and \ref{fig:14}.  It is
interesting to note that the perfect agreement observed in
Fig.~\ref{fig:13} does \textit{not} hold if we reverse the steps, that
is, if we first perform an SED fit at the \textit{Herschel}
resolution, and then convolve the optical-depth map obtain to the
\SI{5}{arcmin} resolution of \textit{Planck}: this is because the
overall fit involves highly nonlinear equations.

\section{Discussion}
\label{sec:discussion}

\begin{figure}[tp!]
  \centering
  \includegraphics[width=\hsize]{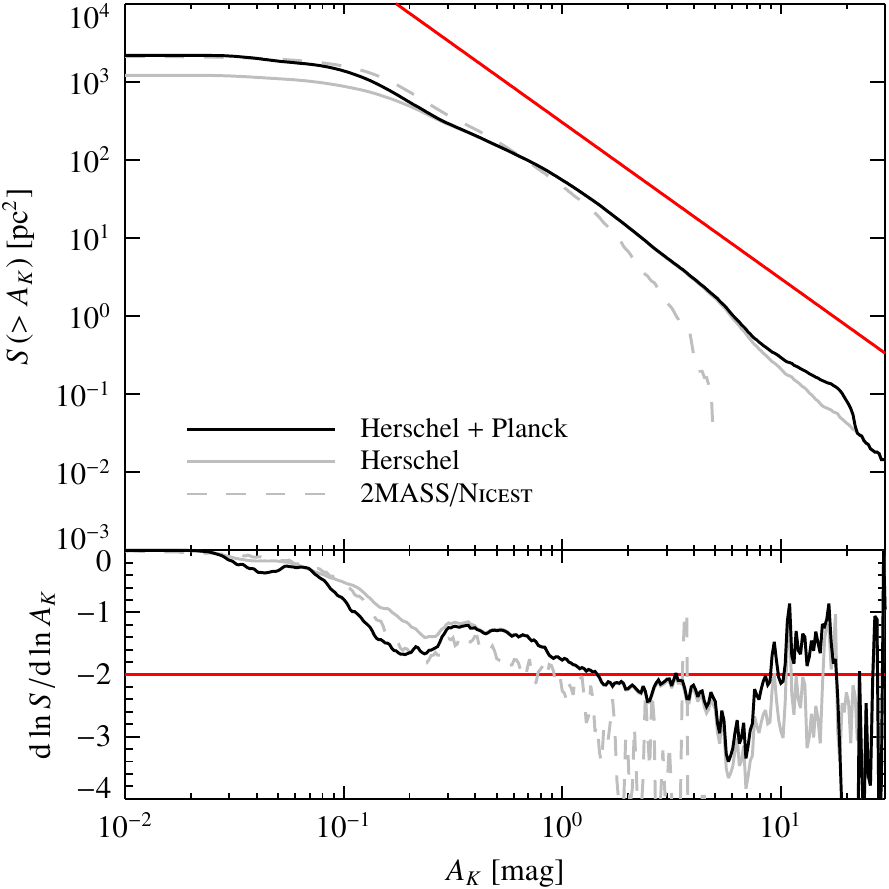}%
  \caption{Integral area-extinction relation for Orion~A, i.e.,
    the physical cloud area above a given extinction threshold (top
    panel), and the logarithmic derivative of this quantity (bottom
    panel).  The solid black line shows the result for the entire
    field, while the solid gray line shows the region covered by
    \textit{Herschel} alone.  For comparison we also plot as a dashed
    gray line the same quantity as obtained from the
    2MASS/\textsc{Nicest} extinction map and a simple $S(>A_K) \propto
    A_K^{-2}$ relation as a red line.}
  \label{fig:15}
\end{figure}

\begin{figure}[tp!]
  \centering
  \includegraphics[width=\hsize]{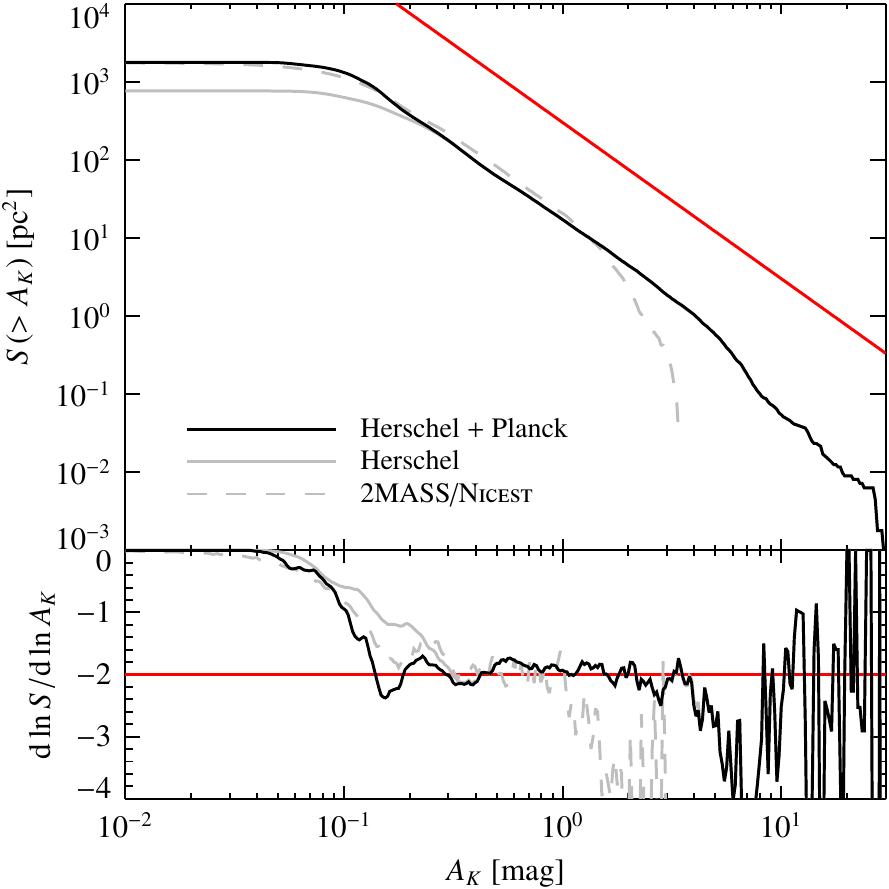}%
    \caption{Same as Fig.~\ref{fig:15} for Orion~B.}
  \label{fig:16}
\end{figure}

\begin{figure}[tp!]
  \centering
  \includegraphics[width=\hsize]{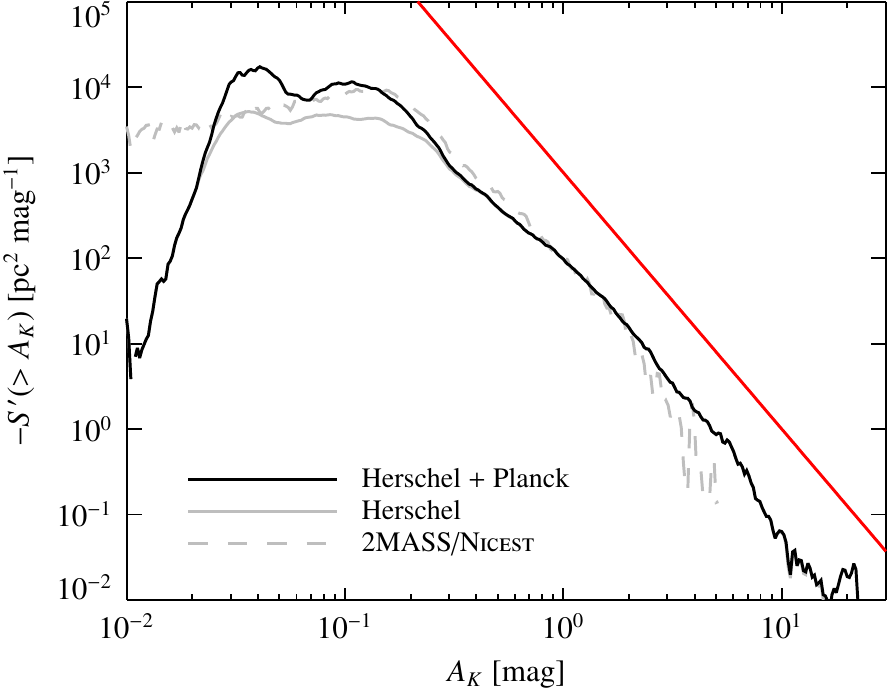}%
  \caption{Function $-S'(>A_K)$, that is, the probability
    distribution function (pdf) of the measured column densities for
    Orion~A.  In this log-log plot a log-normal distribution would appear
    as a parabola, and a power law as a straight line.
    The red line shows the slope of the power law $-S'(A_K) \propto
    A_K^{-3}$.}
  \label{fig:17}
\end{figure}

\begin{figure}[tp!]
  \centering
  \includegraphics[width=\hsize]{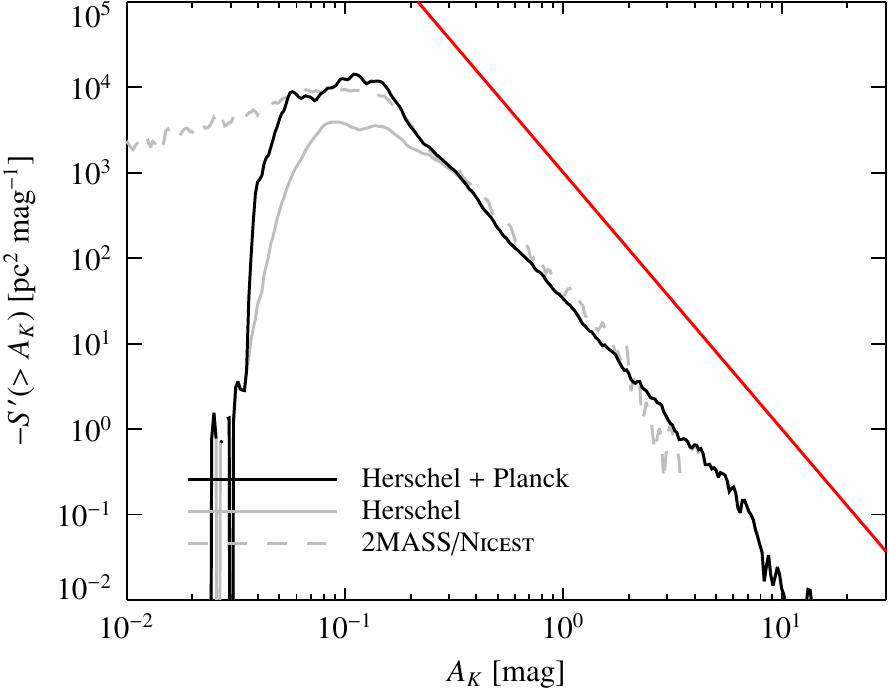}%
  \caption{Same as Fig.~\ref{fig:17} for Orion~B.}
  \label{fig:18}
\end{figure}

The maps we presented can be used for many different purposes, and it
is certainly beyond the scope of this paper to explore all of them in
detail.  Here we just present a few immediate applications, leaving
the rest to follow-up papers.\footnote{The final optical-depth and
  temperature maps presented here are available on the 1 August 2014
  through the website \url{http://www.interstellarclouds.org}.}

Figures~\ref{fig:15} and \ref{fig:16} show the integral area functions
$S(>A_K)$, that is, the cloud area, measured in square parsecs, above
the extinction threshold $A_K$, as a function of $A_K$. The two
figures refer to Orion~A and B, and for both clouds we assumed a
distance of \SI{414}{pc} \citep{2007A&A...474..515M}.  We defined the
boundaries of the clouds to be
\begin{align}
  \label{eq:15}
  &\text{Orion A:} & 206 & {} \le l \le 217 \; , & -21 \le b \le -17
  \; , \notag\\
  &\text{Orion B:} & 203 & {} \le l \le 210 \; , & -17 \le b \le -12
  \; . 
\end{align}
Furthermore, we plot the area functions obtained only in the
\textit{Herschel} covered area (gray solid line), in the total region
that identifies each cloud (black solid line), and as obtained from
the \textsc{Nicest}/2MASS data (gray dashed line).  Note that the gray
solid line is below the black one for low column densities, a result
of the limited area covered by the \textit{Herschel} survey.  At the
other extreme, for high column densities, the dashed line is
consistently below the solid ones, a result of the poorer resolution
and smaller dynamic range of the extinction map derived from the
relatively shallow 2MASS data.  The solid lines are generally
identical in the region covered by the \textit{Herschel} survey (that
is, for $A_K \gtrsim \SI{0.1}{mag}$), except for a small ``bump'' of
the black line in Orion~A for $A_K \sim \SI{10}{mag}$, because of the
lack of \textit{Herschel} data at the center of the OMC due to
saturation (see Fig.~\ref{fig:10}, white area to the left of the
``OMC-1'' label).  Note also that the lowest value for $S(>A_K)$
plotted in Figs.~\ref{fig:15} and \ref{fig:16}, \SI{1D-2}{pc^2},
corresponds to $\sim \SI{8}{pixel}$ in the \SI{36}{arcsec} resolution
optical-depth maps.

For Orion~A, and even more for Orion~B, the $S$ function in a wide
range follows an $A_K^{-2}$ slope, indicated by the red line.  As a
possible explanation of this result, we consider a simple toy model.
The gas in molecular clouds is approximately isothermal, a result of
the combined heating of the gas from cosmic rays and cooling from CO
\citep{1978ApJ...222..881G}.  Therefore, in our toy model it does not
seem unreasonable to use a radial profile $\rho(r) \propto 1/r^2$ for
the dense regions of molecular clouds, corresponding to the singular
isothermal sphere solution (a non-singular isothermal profiles would
follow the $1/r^2$ slope at radii larger than the core).  In
projection, such a density profile would produce a surface density
profile $\Sigma_\mathrm{gas}(r) \propto A_K(r) \propto 1/r$.  Hence,
the cloud \textit{area} above a given extinction threshold in this
simple model would follow the relation $S(> A_K) \propto r^2 \propto
A_K^{-2}$.  Note that this simple argument applies to the gas
component and not to the dust component (which, instead, is easily
heated by starlight).  However, since the gas constitutes the large
majority of mass in the cloud, it is this component that sets the shape
of the $S(> A_K)$ relation; the dust here is merely used as a tracer.

The bottom plots of Figs.~\ref{fig:15} and \ref{fig:16} show the
logarithmic derivative of $S(> A_K)$.  Figure~\ref{fig:16} shows that
the Orion~B cloud is very well described by the $S(>A_K) \propto
A_K^{-2}$ scaling law over nearly two orders of magnitude in
extinction, from $A_K \sim 0.1$ to \SI{5}{mag}.  The $S$ function for
Orion~A (Fig.~\ref{fig:15}) is not as well described by a single
power-law index, but over the same extinction range its slope is quite
close to -2, being somewhat shallower below $A_K \sim \SI{1.0}{mag}$
and somewhat steeper above $A_K \sim \SI{1.0}{mag}$.  At lower column
densities, we observe a clear break of the relation, with the
cumulative area function $S(> A_K)$ approximately constant.  This is
likely due to the limited area considered in making Figs.~\ref{fig:15}
and \ref{fig:16}; however, it is obvious that this scaling law cannot
hold at arbitrarily lower column densities for at least two reasons:
first, one inevitably would be outside the survey area when $A_K$
becomes very small; second, physically it is not plausible that
regions with a very low column-density remain isothermal.  From this
latter point of view, it is intriguing to note that the break at low
extinctions is observed around $A_K \sim \SI{0.1}{mag}$, a value where
most likely the dust shielding just starts to become effective enough
to isolate the gas in the inner part of the molecular clouds from the
interstellar radiation field, thus enabling isothermal density
distributions above this threshold.  Interestingly, this is
approximately the same column-density as for the survival of ${}^{12}$CO.
At the other extreme, the scaling law seems to break for column
densities just below \SI{10}{mag}.  We note, however, that it is not
obvious that this break is real, as it might be a result of systematic
effects that are possibly present at high column densities.  We
mention as a probable source of biases temperature gradients along the
line of sight, which are thus completely unaccounted for in the SED
fit.

Figures~\ref{fig:17} and \ref{fig:18} show the differential area
function, or more precisely, $-S'(> A_K)$.  This function is just
proportional to the probability distribution function of
column-densities (pdf), a function often considered in the context of
molecular cloud studies.  It is generally accepted that this function
has a log-normal shape as a result of the turbulent supersonic motion
that are believed to characterize molecular clouds on large scales
\citep[e.g.][]{1994ApJ...423..681V, 1997ApJ...474..730P,
  1998PhRvE..58.4501P, 1998ApJ...504..835S}.  However, as shown by
\citet{2010MNRAS.408.1089T}, log-normal distributions are also
expected under completely different physical conditions (also
plausible for molecular clouds), such as radially stratified density
distributions dominated by gravity and thermal pressure, or by a
gravitationally driven ambipolar diffusion.  Vice versa, the
log-normality of the pdf has been challenged in clouds located in
isolated environments such as Corona
\citep{2014arXiv1401.2857A}. Surprisingly, our study suggests that the
log-normal regime, if at all present in the cloud studied here, is
confined to very low column densities, below $A_K \sim \SI{0.1}{mag}$;
for higher column densities we again find the scaling-law relation
$S'(> A_K) \propto A_K^{-3}$.  Although this behavior has been
observed in the past, especially in star-forming clouds
\citep{2009A&A...508L..35K}, we are now in the position to demonstrate
that the power-law regime dominates most ranges of column-densities
and generally characterizes the cloud structure above $A_K \sim
\SI{0.1}{mag}$.  We also stress that this limit, corresponding to
\SI{1}{mag} of visual extinction, really marks the boundary of
molecular clouds and is, for instance, also associated to the limit
for the photodissociation of carbon monoxide.

\begin{figure}[tp!]
  \centering
  \includegraphics[width=\hsize]{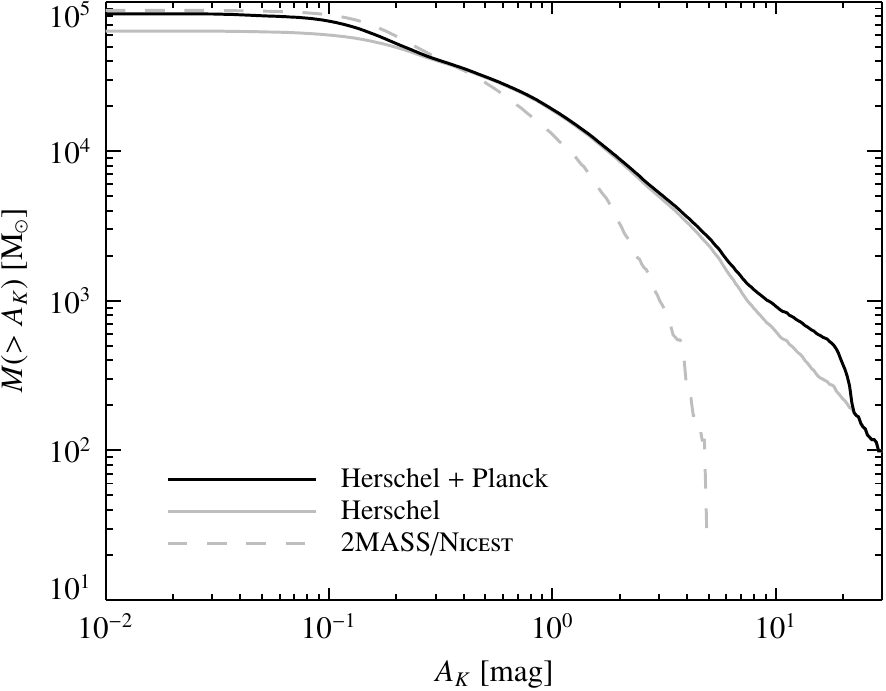}%
  \caption{Integral mass-extinction relation for Orion~A, i.e.\
    the cloud mass above a given extinction threshold.  The color
    codes follow the same convention as in Fig.~\ref{fig:15}.}
  \label{fig:19}
\end{figure}

\begin{figure}[tp!]
  \centering
  \includegraphics[width=\hsize]{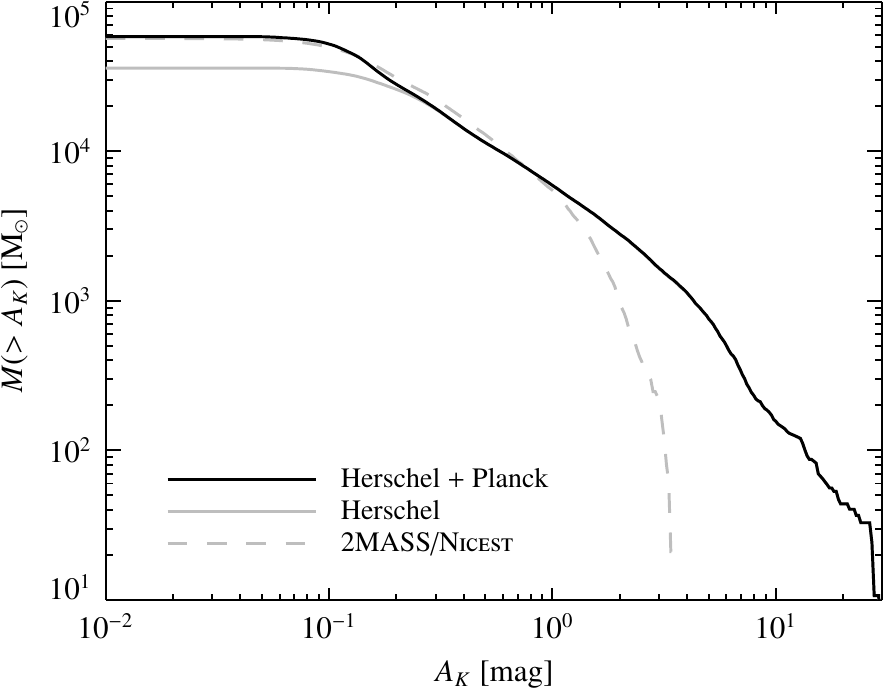}%
    \caption{Same as Fig.~\ref{fig:19} for Orion~B.}
  \label{fig:20}
\end{figure}

Figures~\ref{fig:19} and \ref{fig:20} show the integral mass functions
$M(>A_K)$, that is, the cloud mass above the extinction threshold
$A_K$, as a function of $A_K$.  As before, we plot mass functions
corresponding to the various dataset using different line colors and
styles.  The mass was estimated by converting the column-density into
a surface mass density using the factor
\begin{equation}
  \label{eq:16}
  \frac{\Sigma}{A_K} = \mu \beta_K m_\mathrm{p} \simeq
  \SI{183}{M_{$\odot$}.pc^{-2}} \, ,
\end{equation}
where $\mu \simeq 1.37$ is the mean molecular weight corrected for the
helium abundance, $\beta_K \simeq \SI{1.67D22}{cm^{-2}.mag^{-1}}$ is
the gas-to-dust ratio, that is, $[N(\textsc{Hi}) + 2N(H_2))] / A_K$
(\citealp{1979ARA&A..17...73S, 1955ApJ...121..559L,
  1978ApJ...224..132B}; see also \citealp{1985ApJ...288..618R} for the
conversion from $A_K$ to $A_V$ for 2MASS), and $m_\mathrm{p} =
\SI{1.67D-24}{g}$ is the proton mass.

Note the very good agreement of at low column densities between the
solid black and dashed gray lines, that is, between the ``total''
masses measured from \textit{Herschel} + \textit{Planck} and from
2MASS/\textsc{Nicest}: this clearly is a result of using 2MASS
to calibrate the $\gamma$ factor.

Not unexpectedly, both mass functions of Orion~A and B approximately
follow an $A_K^{-1}$ slope.  This is a direct consequence of the
fact that the area functions follow an $A_K^{-2}$ slope, since we have
\begin{equation}
  \label{eq:17}
  M(>A_K) \propto \int_{A_K}^\infty \frac{\diff S(> A_K')}{\diff A_K'}
  A_K' \, \diff A_K' \; . 
\end{equation}

\begin{figure*}[tp!]
  \centering
  \includegraphics[width=0.49\hsize]{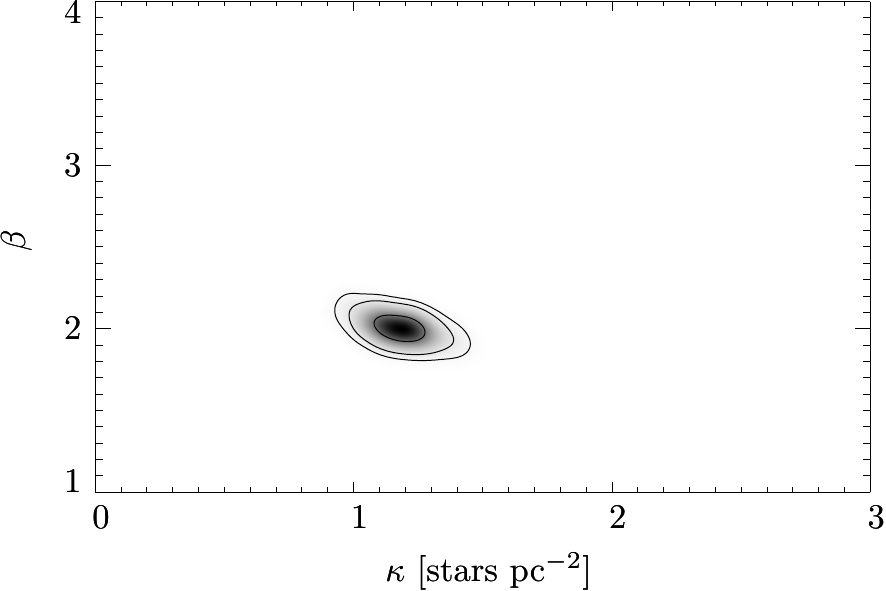}\hfill
  \includegraphics[width=0.49\hsize]{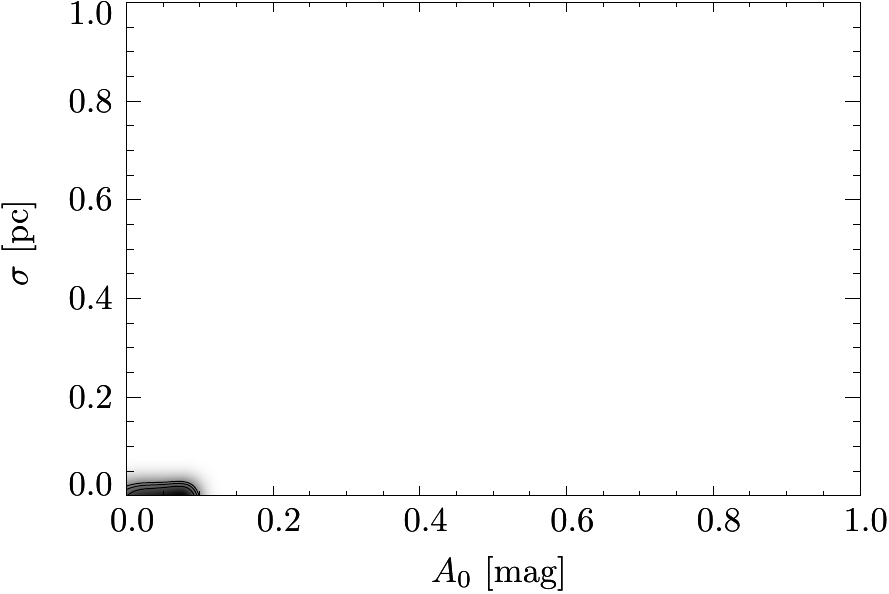}%
  \caption{Posterior probability distributions for the four
    parameters of the model described in Eqs.~\eqref{eq:18} and
    \eqref{eq:19}, obtained with a Markov chain Monte Carlo algorithm
    for Orion~A.  The obtained best-fit values and their formal errors
    are $\kappa = \SI{1.18+-0.09}{stars.pc^{-2}.mag^{-\beta}}$, $\beta
    = \num{1.99+-0.05}$, $A_0 = \SI{0.050+-0.028}{mag}$, and $\sigma =
    \SI{0.0088+-0.0049}{pc}$.}
  \label{fig:21}
\end{figure*}

\begin{figure}[tp!]
  \centering
  \includegraphics[width=\hsize]{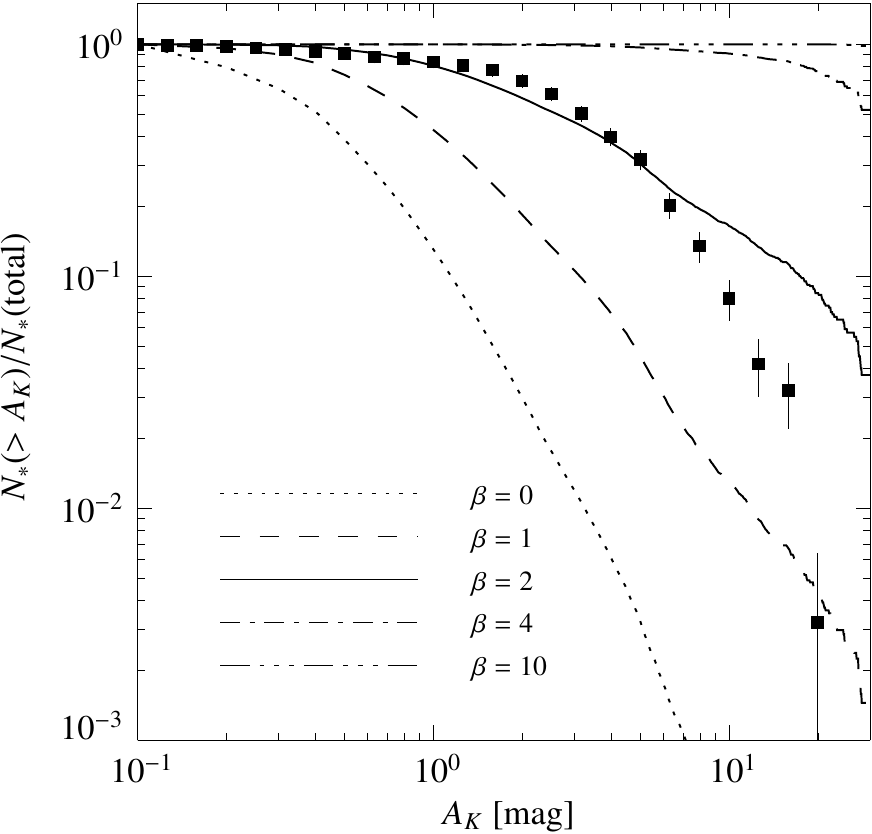}%
  \caption{Fractional number of protostars in Orion~A above a given
    extinction threshold as a function of the threshold, together with
    a few predictions from simple power-law models.  The curve for
    $\beta = 2$ agrees well with the data up to $A_K = \SI{6}{mag}$.}
  \label{fig:22}
\end{figure}

\begin{figure*}[tp!]
  \centering
  \includegraphics[width=0.49\hsize]{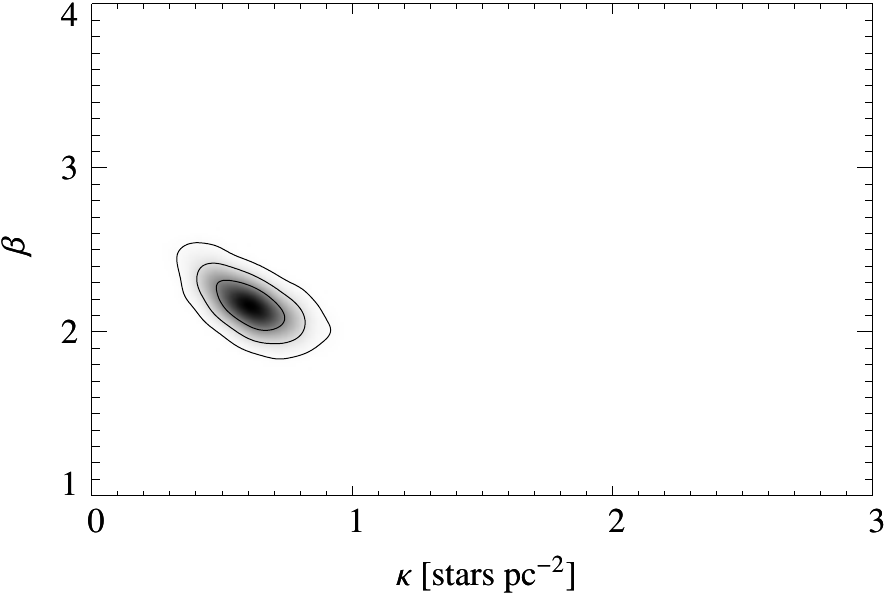}\hfill
  \includegraphics[width=0.49\hsize]{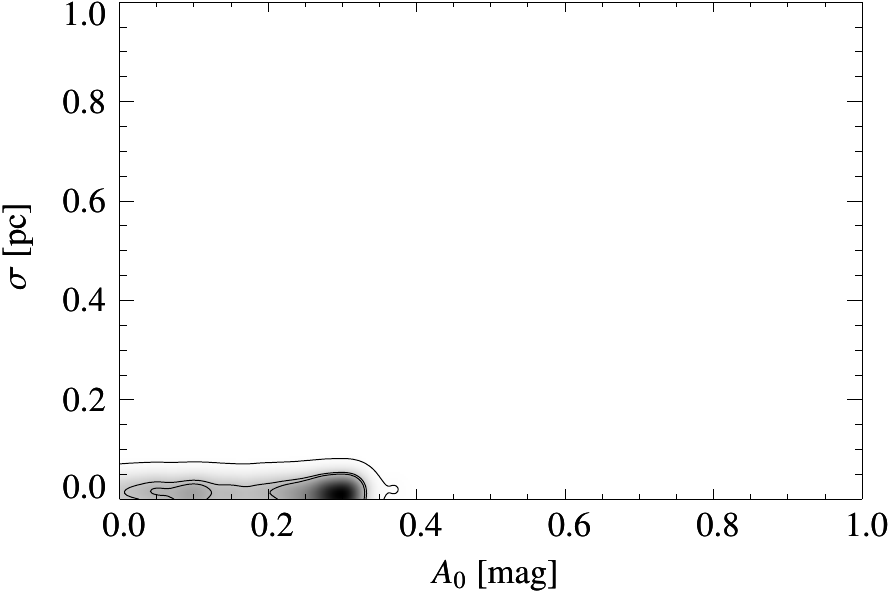}%
  \caption{Posterior-probability distributions for the four
    parameters of the model described in Eqs.~\eqref{eq:18} and
    \eqref{eq:19}, obtained with a Markov chain Monte Carlo algorithm
    for Orion~B.  The obtained best-fit values and their formal errors
    are $\kappa = \SI{0.60+-0.10}{stars.pc^{-2}.mag^{-\beta}}$, $\beta
    = \num{2.16+-0.10}$, $A_0 = \SI{0.20+-0.10}{mag}$, and $\sigma =
    \SI{0.018+-0.007}{pc}$.}
  \label{fig:23}
\end{figure*}

\begin{figure}[tp!]
  \centering
  \includegraphics[width=\hsize]{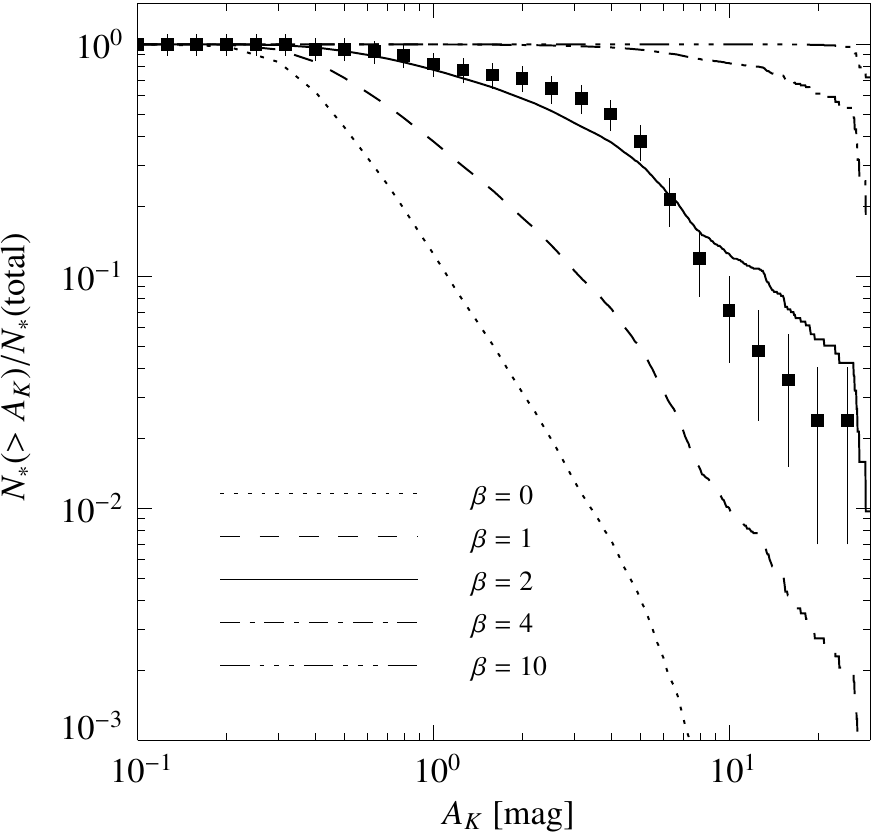}%
  \caption{Fractional number of protostars in Orion~B above a given
    extinction threshold as a function of the threshold, together with
    a few predictions from simple power-law models.  The curve for
    $\beta = 2$ agrees well with the data up to $A_K = \SI{6}{mag}$.}
  \label{fig:24}
\end{figure}

\begin{figure}[tp!]
  \centering
  \includegraphics[width=\hsize]{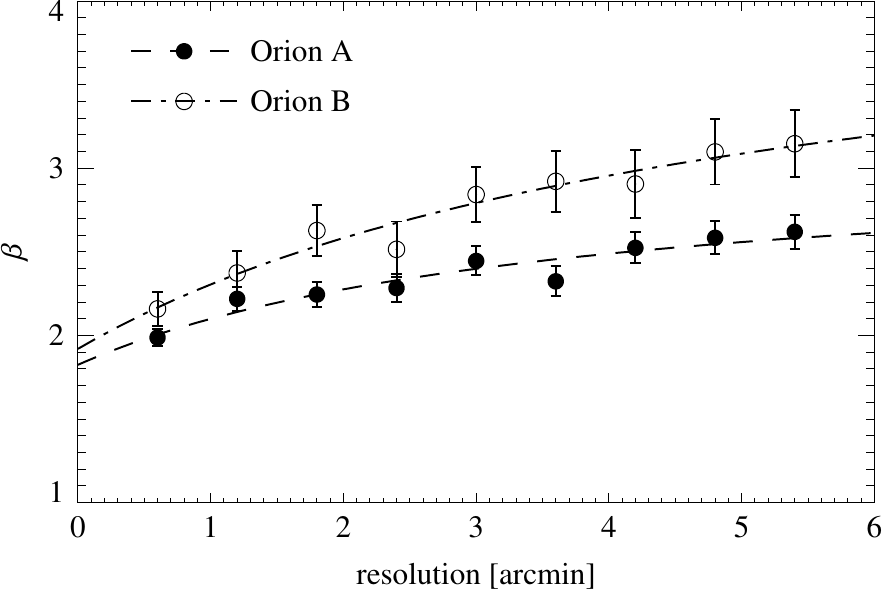}%
  \caption{Behavior of the $\beta$ exponent for increasingly coarser
    resolutions for Orion~A and Orion~B.  The data have been obtained
    by repeating the parameter inference for the model of
    Eqs. \eqref{eq:18} and \eqref{eq:19} on images smoothed with
    different kernel sizes.  The dashed lines show the result of two
    simple rational fits for data of the form $\beta = (\beta_0 + a \,
    \mathit{FWHM}) / (1 + b \, \mathrm{FWHM})$, which seems to
    reproduce the data well.  The best-fit parameters obtained for the
    two clouds are $\beta_0 = \num{1.82+-0.14}$ for Orion~A and
    $\beta_0 = \num{1.92+-0.24}$ for Orion~B.}
  \label{fig:25}
\end{figure}



A possible different application of our maps is to test the validity
of the Kennicutt-Schmidt relation \citep{1959ApJ...129..243S,
  1998ApJ...498..541K}.  This relation refers to globally averaged
surface densities of the star formation rate ($\Sigma_\mathrm{SFR}$)
and the gas ($\Sigma_\mathrm{gas}$) in galaxies.  Recently, we
investigated the \textit{local} form of this relation
\citep{2013A&A...559A..90L} by modeling star formation events in
molecular clouds as a (possibly delayed) Poisson spatial process.  We
showed that the local protostar surface density is simply proportional
to the square of the projected gas density, or equivalently to the
square of the column-density, $A_K$.  In this respect, it is
interesting to consider a simple model of star formation where the
predicted density of protostars at the position $x$,
$\Sigma_\mathrm{YSO}(x)$, is written as the convolution of a
primordial density of protostars $\Sigma^{(0)}_\mathrm{YSO}(x)$ with a
Gaussian kernel:
\begin{equation}
  \label{eq:18}
  \Sigma_\mathrm{YSO}(x) = \int \frac{1}{2 \pi \sigma^2} \e^{|x -
    x'|^2 / 2 \sigma^2} \Sigma_\mathrm{YSO}^\mathrm{(0)}(x') \,
  \diff^2 x' \; .
\end{equation}
The convolution is used to model the fact that the star-formation
process is not instantaneous, and during this process the protostars
might become displace from the original sites of their formation.  Finally,
we model the primordial density of protostars as
\begin{equation}
  \label{eq:19}
  \Sigma_\mathrm{YSO}^\mathrm{(0)}(x) = \kappa \Hs\bigl( A_K(x) - A_0
  \bigr) \left( \frac{A_K}{1 \mbox{ mag}} \right)^\beta(x) \; ,
\end{equation}
where $\Hs$ is the Heaviside function
\begin{equation}
  \label{eq:20}
  \Hs(z) =
  \begin{cases}
    1 & \text{if $z  >  0 \; ,$} \\
    0 & \text{if $z \le 0 \; .$}
  \end{cases}
\end{equation}
Therefore, in our model the constants involved are the normalization
$\kappa$ (taken to be measured in units of
\si{star.pc^{-2}.mag^{-\beta}}), the star formation threshold $A_0$
(in units of $K$-band extinction), the dimensionless exponent $\beta$,
and the diffusion coefficient $\sigma$ (measured in \si{pc}).

Using the technique described in \citet{2013A&A...559A..90L}, we can
use a catalog of protostars and a map of the cloud column-density to
infer the values of the four parameters of the model.  As a sample
application, we show here the results obtained for Orion~A using the
\textit{Spitzer}-based catalog of protostars of
\citet{2012AJ....144..192M} (see \citealp{2013A&A...559A..90L} and
\citealp{2013ApJ...778..133L} for further details on the specific data
used).

The results for Orion~A, reported in Figs.~\ref{fig:21} and
\ref{fig:22}, show that the simple relation $\Sigma_\mathrm{YSO}
\propto A_K^2$ is confirmed to describe the local star formation
process in Orion~A well (more precisely, we find $\beta =
\num{1.99+-0.05}$ within $1$-$\sigma$; but see the figures for
detailed credibility regions).  However, Fig.~\ref{fig:22} also shows
that for $A_K > \SI{6}{mag}$ the observed number of protostars seems
to be below the prediction given by the simple relation
$\Sigma_\mathrm{YSO} \propto A_K^2$.  Several realistic processes
might produce this effect: it might be a genuine result of
evolutionary effects (not all the high-density gas might yet have
produced stars, or stars might have moved from their sites of
formation), or it might be an observational artifact (small-number
statistics or simply our inhability to detect all protostars because
of confusion effects).  Furthermore, the value of $\kappa$ is
significantly below than the value measured by
\citet{2013A&A...559A..90L} using the 2MASS/\textsc{Nicest} map
(\SI{1.18+-0.09}{stars.pc^{-2}.mag^{-\beta}} vs.\
\SI{1.64+-0.09}{stars.pc^{-2}.mag^{-\beta}}).  This is to be expected,
since $\kappa$ is quite sensitive to changes of resolution:
high-resolution maps probe the small peaks of molecular clouds better,
where significant star formation occurs.  In particular, if $\beta >
1$, we expect that a smoothing in the measured map is associated with a
higher measured value of $\kappa$, to compensate for the ``missed''
star-forming density in the high-density peaks.  In contrast, the
value of $\beta$ appears to be more robust (see discussion below).



The results for Orion~B are shown in Figs.~\ref{fig:23} and
\ref{fig:24}.  The parameters obtained in this cloud are consistent
with those of Orion~A; in particular, the simple $\Sigma_\mathrm{YSO}
\propto A_K^2$ relation is verified: we measure $\beta =
\num{2.16+-0.10}$.  The result is somewhat surprising, since in
\citet{2013ApJ...778..133L} we found instead that the star formation
in Orion~B would follow a $\beta \simeq 3$ law.  A closer
investigation shows that this discrepancy can be essentially
attributed to resolution effects.  To prove this assertion,
we repeated the entire local Schmidt-law analysis using maps with
degraded resolution.  The results obtained for both Orion~A and B are
shown in Fig.~\ref{fig:25}: they show that $\beta$ clearly increases
with the final \textit{FWHM} of the images, the effect being limited
for Orion~A, and much more substantial for Orion~B.  Moreover, it
appears that the data to the left of the plot converge around $\beta
\simeq 2$ for both clouds.

Interestingly, in Orion~B we also see a hint for a threshold in the
star-forming rate, that is, it seems probable that $A_0$ is strictly
positive, which is confirming a similar result
\citep{2013ApJ...778..133L}.  We stress, however, that even the current
data cannot exclude the case $A_0 \simeq \SI{0}{mag}$.

In summary, it is intriguing that different clouds seem to be
characterized by essentially the same local Schmidt-law and also show
the same distribution functions of dense gas, with $S(> A_K) \propto
A_K^2$.  These two facts together might be the key to understanding
the good correlation found by \citet{2010ApJ...724..687L} between the
mass of dense gas and the star formation rate in local molecular
clouds.

\section{Conclusions}
\label{sec:conclusions}

Our main results can be summarized in the following items:
\begin{itemize}
\item We presented optical-depth and temperature maps of the entire
  Orion molecular cloud complex obtained from \textit{Herschel} and
  \textit{Planck} space observatories.
\item The maps have a \SI{36}{arcsec} resolution for \textit{Herschel}
  observations and a \SI{5}{arcmin} resolution elsewhere.  In
  addition, we also produced a \SI{18}{arcsec} resolution
  optical-depth maps based on the SPIRE~250 data alone.
\item We calibrated the optical-depth maps using 2MASS/\textsc{Nicest}
  extinction data, thus obtaining column-density extinction maps at
  the resolution of \textit{Herschel} with a dynamic range
  \SI{1D-2}{mag} to \SI{30}{mag} of $A_K$, or from \SI{4D20}{cm^{-2}}
  to \SI{6D23}{cm^{-2}}.
\item We measured $C_{2.2}/\kappa_{850}$, that is, the ratio of the
  \SI{2.2}{\um} extinction coefficient and of the \SI{850}{\um}
  opacity.  We found that the values obtained for both Orion~A and B
  cannot be explained using the \citet{1994A&A...291..943O} or the
  \citet{2001ApJ...548..296W} theoretical models of dust, but agree
  very well with the newer \citet{2011A&A...532A..43O} models for
  ice-covered silicate-graphite conglomerate grains.
\item We examined the cumulative and differential area functions of
  the data, showing that over a large regime of extinction we observe
  a power-law $S(> A_K) \propto A_K^{-2}$, which is reminiscent of a
  simple isothermal model of molecular clouds; surprisingly, we do not
  see clear evidence of log-normality in the column-density pdf.
\item We used the \textit{Planck/Herschel} maps to re-evaluate the
  \textit{local} Schmidt-law for star formation, $\Sigma_{YSO}
  \propto A_K^\beta$.  We found that $\beta \simeq 2$ in Orion~A,
  confirming our earlier studies \citep{2013A&A...559A..90L,
    2013ApJ...778..133L}.  For Orion~B, we also found $\beta \simeq
  2$, which is lower than our previous estimates as a result of the
  much improved angular resolution of the \textit{Herschel}
  observations.
\end{itemize}

\begin{acknowledgements}
  Based on observations obtained with \textit{Planck}
  (\url{http://www.esa.int/Planck}), an ESA science mission with
  instruments and contributions directly funded by ESA Member States,
  NASA, and Canada. We are grateful to H.~Roussel for her help with
  Scanamorphos. H.~Bouy is funded by the Ram\'on y Cajal fellowship
  program number RYC-2009-04497. J.~Alves acknowledges support from
  the Faculty of the European Space Astronomy Centre (ESAC).
\end{acknowledgements}

\appendix

\section{Hidden layers of multiple-layer figures}

In this appendix we provide a ``flat'' version of the hidden layers of
multi-layer figures, useful if no JavaScript-enabled PDF reader is
used, or for the printed version of the paper.

\begin{figure}[h!]
  \centering
  \includegraphics[width=\hsize]{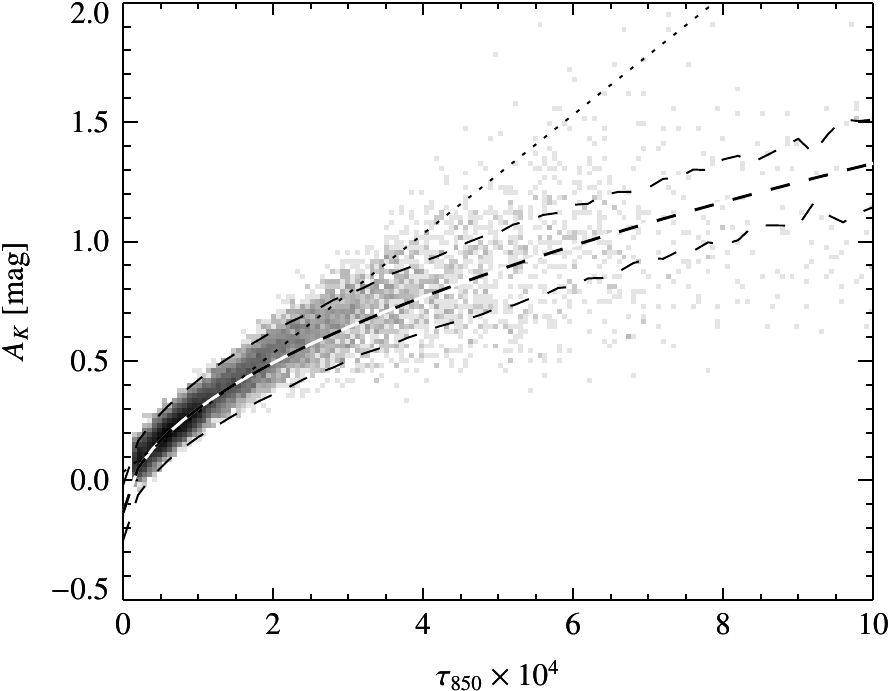}%
  \caption{Relationship between submillimiter optical-depth and
    \textsc{Nicer} extinction map in Orion~A.  The plot shows clear
    nonlinear effects for high values of dust column-densities or
    optical-depths, as shown by the curved fit (dashed line).  We also
    report in this plot the linear fit obtained in the range of
    Fig.~\ref{fig:3} (dotted line).  The same figure with
    \textsc{Nicest} technique is provided in Fig.~\ref{fig:5}.}
  \label{fig:105}
\end{figure}

\begin{figure}[h!]
  \centering
  \includegraphics[width=\hsize]{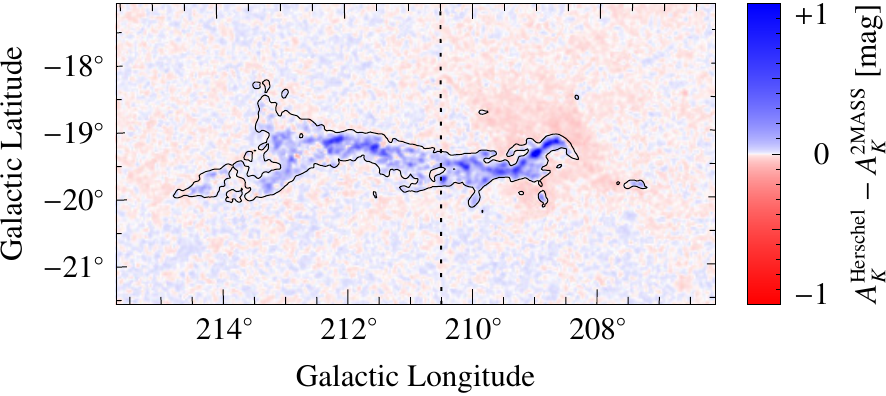}%
  \caption{Difference between the extinction predicted by
    \textit{Herschel}, using Eq.~\eqref{eq:11}, and the extinction
    measured with 2MASS/\textsc{Nicer}, with blue (red)
    indicating a positive (negative) difference.  The same plot for
    \textsc{Nicest} is available in Fig.~\ref{fig:6}.}
  \label{fig:106}
\end{figure}

\begin{figure*}[h!]
  \centering
  \includegraphics[width=\hsize]{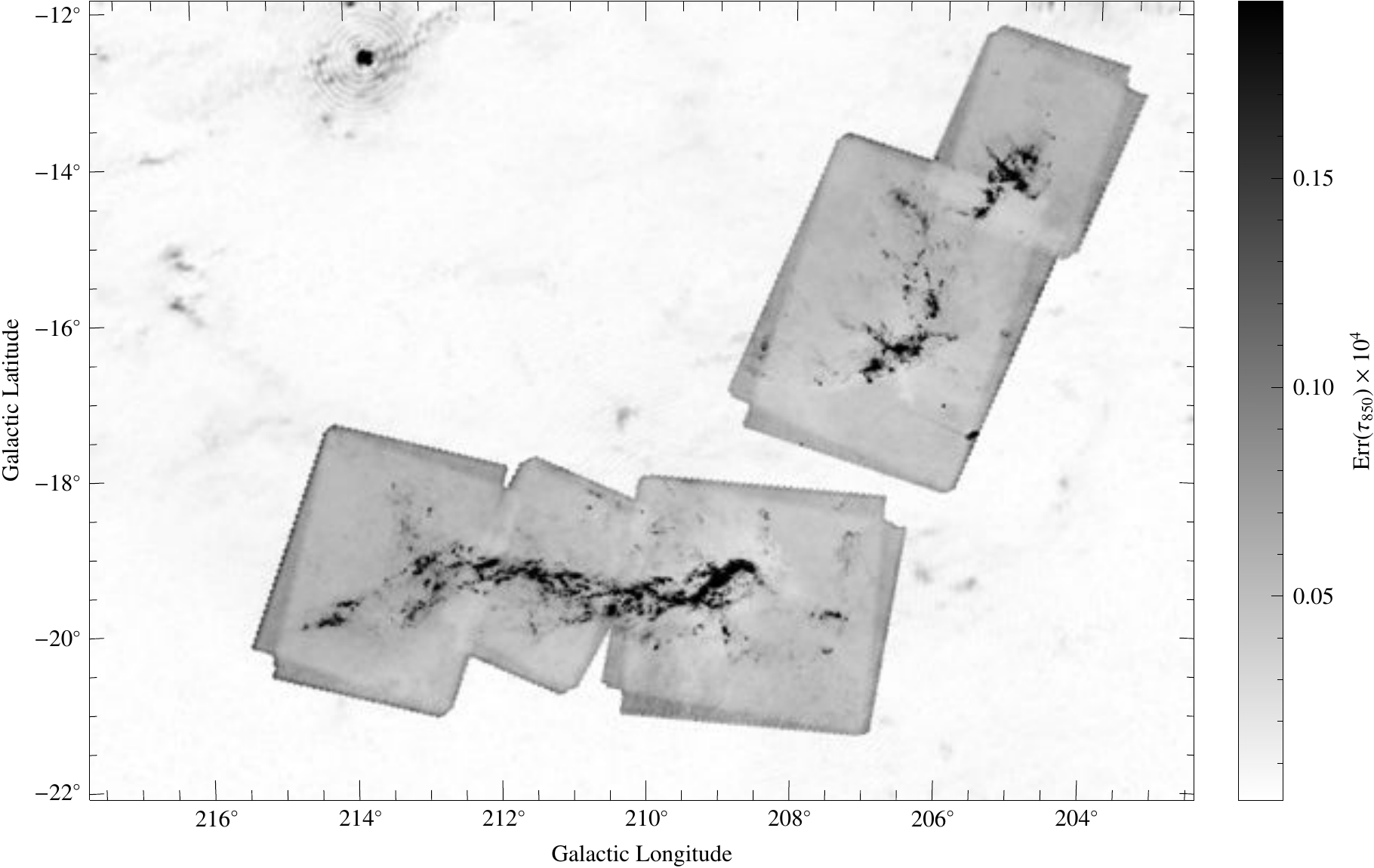}%
  \caption{Error on the optical-depth for the image reported in
    Fig.~\ref{fig:8}.  The figure clearly shows the areas where the
    \textit{Herschel} data are available.  The resolution of the image
    varies from \SI{5}{arcmin} (corresponding to the \textit{Planck}
    data) to \SI{36}{arcsec} (for the \textit{Herschel}-covered
    areas). }
  \label{fig:108}
\end{figure*}

\begin{figure*}[h!]
  \centering
  \includegraphics[width=\hsize]{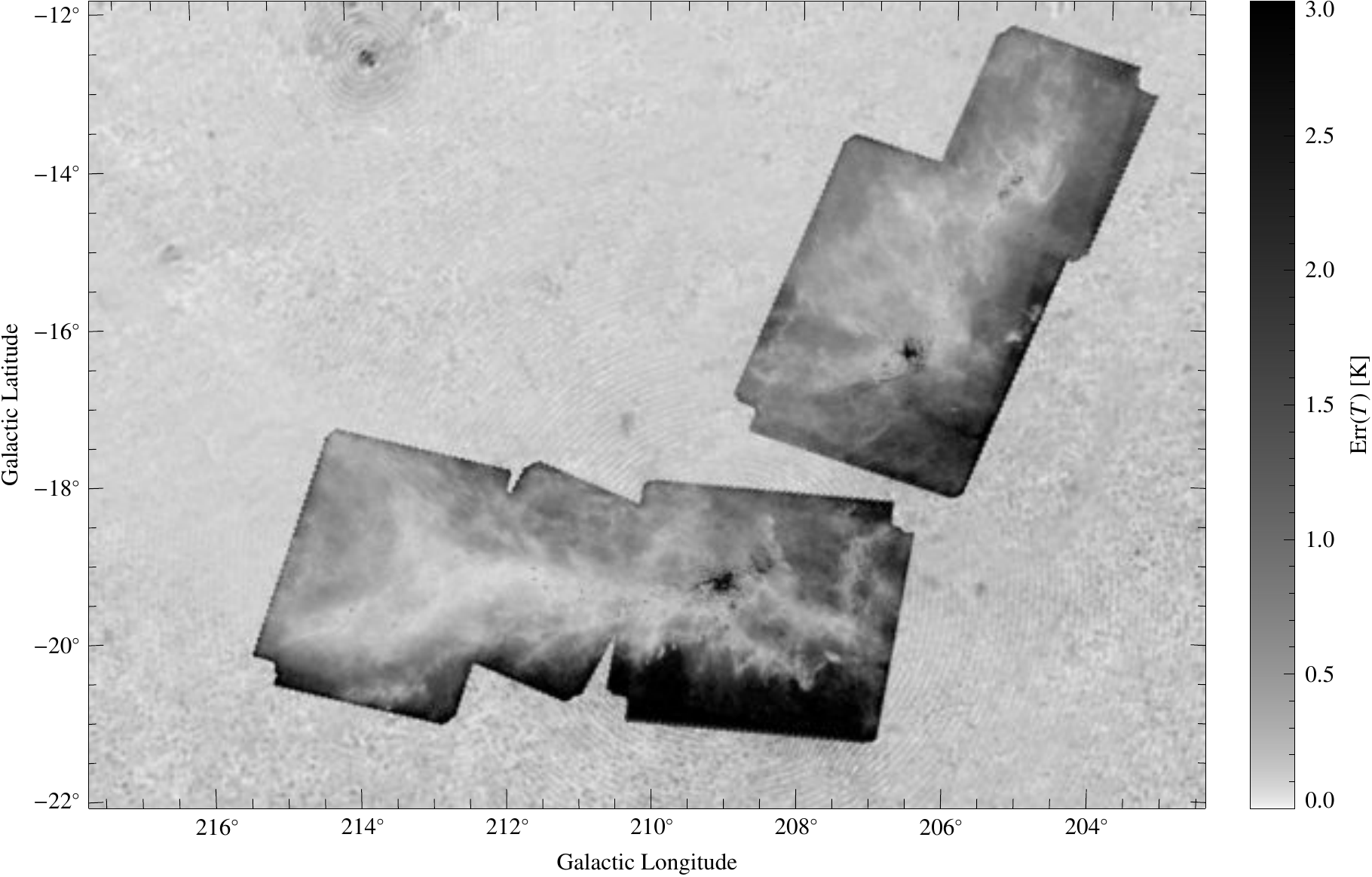}%
  \caption{Error on the effective dust temperature for the image
    reported in Fig.~\ref{fig:9}.}
  \label{fig:109}
\end{figure*}

\bibliographystyle{aa} 
\bibliography{../dark-refs}

\end{document}